\documentclass[aps,prd,twocolumn,superscriptaddress,nofootinbib]{revtex4-2}

\usepackage{supertabular} 
\usepackage{placeins}
\usepackage{epsfig}
\usepackage{graphicx}

\usepackage{float}
\usepackage{mathtools}
\usepackage{array}
\usepackage{dcolumn}

\usepackage{latexsym}
\usepackage{amsmath}
\usepackage{amssymb}
\usepackage{amsfonts}
\usepackage{bm}

\usepackage[T1]{fontenc}
\usepackage{lmodern}
\usepackage{booktabs}

\usepackage{color}
\definecolor{purple}{rgb}{0.5,0,0.5}
\definecolor{blue}{rgb}{0.0,0,0.9}
\definecolor{prdblue}{rgb}{0.133,0.118,0.498}
\usepackage[colorlinks=true, pdfstartview=FitV, linkcolor=prdblue, citecolor= prdblue, urlcolor=prdblue]{hyperref}

\newcommand{\dd}{\mathrm{d}}

\hypersetup{pdftitle={From electromagnetic form factors to proton generalized parton distributions},pdfsubject={Pedagogical PRD working manuscript; diagnostic benchmark},pdfkeywords={GPD, proton, electromagnetic form factors, double distributions, D-term}}

\def\tstrut{\vrule height3.25ex depth0pt width0pt} 

\begin{document}


\title{Proton generalized parton distributions from electromagnetic form factors: A theoretically constrained phenomenological reconstruction}

\author{A.~Capelo-Astudillo}
\email{ancapelo@espol.edu.ec}
\affiliation{Escuela Superior Politécnica del Litoral, Facultad de Ciencias Naturales y Matemáticas, Guayaquil, Ecuador}

 \author{T.~Aguilar}
\affiliation{Departamento de Física, Colegio de Ciencias e Ingeniería, Universidad San Francisco de Quito, Quito 17901, Ecuador}

\author{M.~Conde-Correa}
\affiliation{Departamento de Física, Colegio de Ciencias e Ingeniería, Universidad San Francisco de Quito, Quito 17901, Ecuador}

\author{A.~Duenas-Vidal}
\affiliation{Departamento de Física, Escuela Politécnica Nacional, Quito 170143, Ecuador}

\author{P.~G.~Ortega}
\affiliation{Departamento de F\'isica Fundamental, Universidad de Salamanca, E-37008 Salamanca, Spain}
\affiliation{Instituto Universitario de F\'isica 
Fundamental y Matem\'aticas (IUFFyM), Universidad de Salamanca, E-37008 Salamanca, Spain}

\author{J.~Segovia}
\email[]{jsegovia@upo.es}
\affiliation{Departamento de Sistemas F\'\i sicos, Qu\'\i micos y Naturales, Universidad Pablo de Olavide, Avenida Rectora Rosario Valpuesta 1, 41089, Dos Hermanas, Sevilla, Spain.}

\date{\today}

\begin{abstract}
We present a theoretically constrained phenomenological reconstruction of the unpolarized, chiral-even, leading-twist proton generalized parton distributions $H^q$ and $E^q$. At zero skewness, longitudinal valence-quark distributions are combined with flavor-dependent exponential profiles that correlate the longitudinal momentum fraction with the momentum-transfer dependence. The profile parameters and the forward limit of $E^q$ are calibrated simultaneously against phenomenological representations of the proton and neutron electromagnetic form factors. Exact valence and anomalous-magnetic-moment normalizations, the low-scale quark-sector condition $B^{u+d}(0)=0$, the required endpoint behavior, and the complete Gaussian impact-parameter positivity bound are imposed throughout the calibration. The resulting parametrization provides a reasonable overall description of the electromagnetic form factors and their slopes, while exposing the tension between strict GPD consistency and a precision reproduction of all low-$Q^2$ electromagnetic observables within the adopted ansatz.
The zero-skewness distributions are extended to nonzero skewness using normalized double distributions, whereas an independent $D$-term supplies the highest powers allowed in the relevant Mellin polynomials. We derive explicitly the support and integration domains, electromagnetic sum rules, low Mellin moments, impact-parameter distributions, positivity conditions, and the connection with energy--momentum-tensor form factors. The analysis separates the information inherited from the electromagnetic calibration from the additional assumptions governing skewness and mechanical properties. It thereby illustrates both the predictive content and the intrinsic model dependence of reconstructing GPDs from their lowest moments. The resulting framework provides a transparent and  systematically testable elastic baseline for subsequent studies of nucleon-to-resonance transition GPDs.
\end{abstract}

\maketitle


\section{Introduction}
\label{sec:introduction}

Understanding how the mass, spin, and spatial structure of the proton emerge from its quark and gluon constituents remains a central problem in hadron physics. Different experimental processes provide complementary projections of this structure. Inclusive deep-inelastic scattering constrains parton distribution functions (PDFs), which describe the longitudinal momentum carried by partons in a fast-moving hadron. Elastic lepton--nucleon scattering gives access to electromagnetic form factors, which characterize the response of the nucleon to an external electromagnetic current as a function of the invariant momentum transfer. Neither class of observables, considered separately, determines how the longitudinal motion of partons is correlated with their transverse spatial distribution.

Generalized parton distributions (GPDs) provide a unified framework for establishing this connection~\cite{Mueller1994, JiDVCS1997, Radyushkin1996, Diehl2003, Belitsky2005}. They are defined through nonforward matrix elements of quark and gluon light-ray operators and depend on the average longitudinal momentum fraction $x$, the longitudinal momentum asymmetry or skewness $\xi$, the invariant momentum transfer $t$, and the factorization scale $\mu$. At leading twist, GPDs enter the QCD factorization of hard exclusive reactions such as deeply virtual Compton scattering (DVCS) and deeply virtual meson production (DVMP)~\cite{JiDVCS1997, CollinsFrankfurtStrikman1997, BelitskyMullerKirchner2002}. Measurements at Jefferson Lab and, in the future, at the Electron--Ion Collider are expected to improve substantially our knowledge of the multidimensional partonic structure of hadrons~\cite{EICYellowReport2022}.

The physical content of GPDs is encoded in several complementary limits and moments. In the forward limit, the helicity-conserving quark GPD $H^q$ reduces to the ordinary quark and antiquark distributions. The first Mellin moments of $H^q$ and $E^q$ yield the flavor-separated Dirac and Pauli electromagnetic form factors, whereas their second moments are related to form factors of the energy--momentum tensor (EMT). In particular, the combination of the spin-two form factors $A^q$ and $B^q$ enters Ji's angular-momentum relation~\cite{Ji1997}. At zero skewness, a Fourier transform with respect to the transverse momentum transfer produces impact-parameter distributions, thereby correlating the longitudinal momentum fraction of a parton with its transverse position~\cite{Burkardt2000, Diehl2002}. The EMT form factors also enter spatial representations of energy and mechanical properties, including pressure and shear-force distributions~\cite{PolyakovSchweitzer2018, Burkert2018}. Such mechanical interpretations require a specified localization prescription and, for an equilibrium interpretation, the complete conserved EMT. These relations illustrate the broad range of hadron-structure information encoded in GPDs.

This richness is accompanied by an important limitation. GPDs depend on several variables, whereas experimental observables generally probe only weighted integrals or convolutions of them. Electromagnetic form factors constrain their lowest moments in $x$, while deeply virtual observables involve convolution integrals with perturbatively calculable coefficient functions. Reconstructing a GPD from elastic or hard-exclusive data is therefore an inverse problem without a unique solution unless assumptions are introduced about its functional form, flavor dependence, factorization scale, momentum-transfer dependence, and skewness dependence. Polynomiality, support, discrete symmetries, endpoint behavior, and positivity substantially restrict the allowed functions, but they do not eliminate the intrinsic functional freedom. A phenomenological reconstruction must consequently distinguish information constrained by data from information inherited from the longitudinal input and from assumptions introduced through the chosen ansatz.

An economical starting point is the zero-skewness sector. A specified longitudinal distribution can be combined with an $x$-dependent transverse profile whose parameters are constrained by the nucleon electromagnetic form factors. This strategy has been applied successfully in phenomenological analyses of the proton and neutron~\cite{Diehl2005, DiehlKroll2013, Hashamipour2023}; broader reviews of GPD phenomenology emphasize both its usefulness and its unavoidable parametrization ambiguities~\cite{Diehl2003, Belitsky2005, Guidal2013}. At $\xi=0$, this construction gives direct access to correlations between longitudinal momentum and transverse localization and provides a natural representation of the flavor structure of the Dirac and Pauli form factors. Nevertheless, the form-factor sum rules constrain only integrals over $x$. Different GPDs can reproduce similar electromagnetic form factors while differing substantially in their higher moments, impact-parameter distributions, and extrapolations outside the calibrated momentum-transfer interval.

Likewise, correct charge and anomalous-magnetic-moment normalizations do not guarantee a consistent angular-momentum decomposition. In the valence-only description adopted here, imposing $B^{u+d}(0)=0$ assumes that the modeled $u$- and $d$-quark sector saturates the anomalous-gravitomagnetic-moment sum rule at the hadronic input scale. This is an explicit matching condition: in full QCD, the exact relation involves the sum over all active quark flavors and the gluon contribution.

Extending a zero-skewness parametrization to $\xi\neq0$ introduces additional functional freedom. A convenient construction is the double-distribution representation, which implements the correct spectral support and generates polynomial Mellin moments~\cite{Radyushkin1999}. The profile controlling the width of the double distribution, and hence the resulting skewness dependence, is not fixed by electromagnetic form factors. Furthermore, the factorized double-distribution ansatz adopted in this work does not generate the highest power of $\xi$ allowed in the relevant Mellin polynomials. We supply this contribution through an independent $D$-term~\cite{PolyakovWeiss1999}. Its magnitude, flavor decomposition, scale dependence, and momentum-transfer dependence cannot be deduced from the electromagnetic calibration. It is therefore essential to keep the zero-skewness calibration, the double-distribution profile, and the mechanical input conceptually distinct.

The purpose of the present work is to construct a transparent and systematically testable phenomenological reconstruction of the unpolarized, chiral-even, leading-twist proton GPDs $H^q$ and $E^q$. We begin from valence-quark distributions at a hadronic input scale and employ flavor-dependent exponential profiles at zero skewness. Their parameters are calibrated simultaneously against phenomenological representations of the proton and neutron electromagnetic form factors. The forward limit of $E^q$ is allowed to differ in shape from the corresponding valence PDF through a normalization-preserving deformation. During the calibration, we impose exact valence and anomalous-magnetic-moment normalizations, the low-scale condition $B^{u+d}(0)=0$, the required endpoint behavior, and the complete impact-parameter positivity bound associated with the adopted Gaussian ansatz.

The simultaneous treatment of proton and neutron observables is essential for flavor separation. In particular, the neutron form factors involve significant cancellations between the $u$- and $d$-quark contributions and are consequently sensitive to comparatively small changes in the longitudinal input or calibration strategy. The resulting optimum is required to satisfy the theoretical GPD constraints even when this prevents a precision reproduction of every electromagnetic observable within the restricted functional ansatz. The corresponding deviations, especially in quantities controlled by the low-$Q^2$ slopes, provide information about the limitations of the parametrization rather than justification for relaxing the underlying consistency conditions.

We subsequently extend the zero-skewness distributions to nonzero $\xi$ using normalized double distributions and introduce the $D$-term as an independent contribution. Particular attention is paid to the derivation of the integration domain and to analytic and numerical checks of support, normalization, and polynomiality. We also examine the resulting electromagnetic radii, impact-parameter densities, angular-momentum combinations, and EMT form factors. Finally, an analytic tripole model for an independently specified total $D$-form factor is used to illustrate the connection with static pressure and shear-force distributions. This mechanical example is not inferred from the electromagnetic calibration.

Throughout the analysis, we distinguish calibration to phenomenological representations of elastic form factors from a direct likelihood analysis of the underlying experimental measurements. The parameter variations obtained from the adopted calibration therefore characterize sensitivity within the chosen parametrization and objective function; they should not be interpreted automatically as experimental confidence intervals. This distinction is necessary for assessing both the predictive content and the limitations of the reconstructed GPDs.

A further motivation is to establish the elastic reference needed for future studies of nucleon-to-resonance transition GPDs. Transition distributions encode interference between different baryon states and do not, in general, admit the ordinary density interpretation available in the elastic zero-skewness limit. Unequal initial and final masses and different spin-parity assignments modify the covariant decomposition and the relations between Mellin moments and transition form factors, while the finite widths of resonances introduce additional conceptual and phenomenological complications~\cite{Transition2025}. An elastic ansatz therefore cannot simply be relabeled as a transition GPD. The proton case nevertheless provides the necessary testing ground for identifying which ingredients follow from general field-theoretical constraints, which are constrained by form factors, and which remain model dependent.

The remainder of this article is organized as follows. In Sec.~\ref{sec:definitions}, we define the kinematics, operator conventions, support regions, and flavor combinations. The zero-skewness ansatz and its endpoint behavior are introduced in Sec.~\ref{sec:zero-skewness}. Section~\ref{sec:electromagnetic} describes the electromagnetic calibration, normalization checks, and radii, while Sec.~\ref{sec:impact} discusses impact-parameter distributions and positivity. The EMT form factors and angular-momentum relations are examined in Sec.~\ref{sec:gravitational}. The double-distribution extension and the independent $D$-term are developed in Secs.~\ref{sec:nonzero-skewness} and~\ref{sec:D-term}, respectively. Mechanical distributions and their Fourier-transform conventions are considered in Sec.~\ref{sec:mechanical-distributions}. Finally, Sec.~\ref{sec:conclusions} contains our conclusions and outlook.


\section{Definitions, kinematics, and flavor structure}
\label{sec:definitions}

\subsection{Light-front matrix element and elastic kinematics}
\label{sec:light-front-definition}

Let $p$ and $p'$ denote the initial and final proton momenta, respectively, with
\begin{equation}
p^2 = p'^2 = M^2 .
\end{equation}
We introduce the average momentum, momentum transfer, and invariant momentum transfer as
\begin{equation}
P=\frac{p'+p}{2}, \qquad \Delta=p'-p, \qquad t=\Delta^2 .
\label{eq:elastic-kinematics}
\end{equation}
Light-front components are defined by $v^\pm=(v^0\pm v^3)/\sqrt{2}$, and we use the metric $g^{\mu\nu}=\mathrm{diag}(1,-1,-1,-1)$. The skewness parameter is
\begin{equation}
\xi=-\frac{\Delta^+}{2P^+}.
\label{eq:skewness-definition}
\end{equation}
Accordingly, the incoming and outgoing proton plus momenta are $(1+\xi)P^+$ and $(1-\xi)P^+$, respectively. In the quark-emission region, the active quark carries plus momenta $(x+\xi)P^+$ before and $(x-\xi)P^+$ after the hard interaction.

For a quark of flavor $q$, the unpolarized light-front correlator is
defined at $z^+=0$ and $\boldsymbol{z}_\perp=0$ by
\begin{align}
F^q_{s's}(x,\xi,t;\mu) 
&=
\frac{1}{2} \int\frac{dz^-}{2\pi}\, e^{ixP^+z^-} \nonumber \\
&
\times \langle p',s'| \overline q(-z/2)\gamma^+ W[-z/2,z/2]q(z/2) |p,s\rangle \nonumber\\
&=
\frac{1}{2P^+} \overline u(p',s') \Bigg[ \gamma^+ H^q(x,\xi,t;\mu) \nonumber \\
&+ 
\frac{i\sigma^{+\alpha}\Delta_\alpha}{2M} E^q(x,\xi,t;\mu) \Bigg] u(p,s).
\label{eq:gpd-definition}
\end{align}
Here $W[-z/2,z/2]$ is the straight Wilson line along the lightlike separation, and the spinors are normalized according to $\overline u(p,s)u(p,s)=2M$. The scale $\mu$ specifies the renormalization of the nonlocal light-ray operator. It should not be confused with the virtuality of the electromagnetic probe, although the factorization and renormalization scales are often chosen of the same order in hard-scattering applications. Equation~\eqref{eq:gpd-definition} fixes our conventions for $H^q$ and $E^q$~\cite{Diehl2003, Belitsky2005}.

The frequently used statement that $H^q$ conserves proton helicity whereas $E^q$ flips it is most transparent at zero skewness. At $\xi\neq0$, physical proton-helicity amplitudes generally involve kinematic combinations of both distributions. It is therefore the covariant decomposition in Eq.~\eqref{eq:gpd-definition}, rather than the helicity shorthand, that defines the two GPDs.

In a symmetric frame, elastic kinematics gives
\begin{equation}
t = -\frac{\boldsymbol{\Delta}_\perp^2+4\xi^2M^2}{1-\xi^2},
\qquad
t_0 = -\frac{4\xi^2M^2}{1-\xi^2},
\label{eq:t-xi-relation}
\end{equation}
where $t_0$ is the minimum value of $-t$ at fixed $\xi$, or, equivalently, the largest kinematically allowed value of $t$. A physical elastic point satisfies $t\leq t_0$. At fixed spacelike $t<0$, this condition implies
\begin{equation}
|\xi| \leq \sqrt{\frac{-t}{4M^2-t}}.
\label{eq:xi-bound}
\end{equation}
For example, $t=-0.5~\mathrm{GeV}^2$ permits $|\xi|\lesssim0.353$, so that $\xi=0.3$ is kinematically accessible. Notice that $t$ is not a purely transverse momentum transfer when $\xi\neq0$.

\subsection{Support, antiquarks, and valence combinations}
\label{sec:support-valence}

Quark GPDs have support in the interval $-1\leq x\leq1$. Their forward limit obeys
\begin{equation}
H^q(x,0,0;\mu)
=
\begin{cases}
q(x;\mu), & x>0,\\[2mm]
-\overline q(-x;\mu), & x<0.
\end{cases}
\label{eq:gpd-forward-limit}
\end{equation}
No analogous ordinary PDF interpretation exists for $E^q(x,0,0;\mu)$, although its first moment is fixed by the Pauli form factor.

Hermiticity and time-reversal invariance make the elastic GPDs $H^q$ and $E^q$ even functions of $\xi$. We may therefore restrict the discussion to $\xi\geq0$. In that convention, the regions $x>\xi$ and $x<-\xi$ correspond to quark and antiquark Dokshitzer--Gribov--Lipatov--Altarelli--Parisi (DGLAP) regions, respectively, whereas $|x|<\xi$ is the Efremov--Radyushkin--Brodsky--Lepage (ERBL) region, associated with the emission or absorption of a quark--antiquark pair. The negative-$x$ part of the ERBL region should not be interpreted as an ordinary antiquark density.

At positive $x$ and zero skewness, we define the valence
combinations
\begin{align}
H_v^q(x,t)
&=
H^q(x,0,t)+H^q(-x,0,t),
\label{eq:Hv-definition}\\
E_v^q(x,t)
&=
E^q(x,0,t)+E^q(-x,0,t).
\label{eq:Ev-definition}
\end{align}
Equation \eqref{eq:gpd-forward-limit} then gives
\begin{equation}
H_v^q(x,0)=q(x)-\overline q(x) \equiv q_v(x).
\label{eq:Hv-forward}
\end{equation}
In the low-scale construction adopted below, we retain only positive-$x$ dressed-valence degrees of freedom and set the forward antiquark component to zero. Within this boundary condition, $q(x)=q_v(x)$ for $x>0$. This equality is a property of the adopted low-scale model and does not hold at a generic perturbative factorization scale.

The difference between valence and quark-plus-antiquark combinations is especially important when comparing the first and second Mellin moments. Indeed, splitting the complete support interval at $x=0$, using the negative-$x$ convention in Eq.~\eqref{eq:gpd-forward-limit}, and changing variables to $y=-x$ in the negative-$x$ contribution gives the general $(n+1)$th Mellin moment,
\begin{equation}
\int_{-1}^{1}\dd x\,x^n H^q(x,0,0) = \int_0^1\dd y\,y^n \left[ q(y)+(-1)^{n+1}\overline q(y) \right].
\label{eq:full-forward-moment}
\end{equation}
The antiquark contribution therefore enters with a minus sign in the first Mellin moment, corresponding to $n=0$, and with a plus sign in the second Mellin moment, corresponding to $n=1$. Consequently, at $t=\xi=0$,
\begin{align}
\int_{-1}^{1}dx\,H^q(x,0,0) &= \int_0^1dx\, \left[q(x)-\overline q(x)\right],
\label{eq:first-forward-moment}\\
\int_{-1}^{1}dx\,xH^q(x,0,0) &= \int_0^1dx\,x \left[q(x)+\overline q(x)\right].
\label{eq:second-forward-moment}
\end{align}
Thus, the electromagnetic first moment is sensitive to a valence combination, whereas the momentum fraction involves the sum of quark and antiquark distributions. A valence GPD constrained by elastic form factors does not, by itself, determine the complete quark momentum fraction at an arbitrary scale.

More generally, polynomiality gives the spin-two moments in the conventions used later in this work:
\begin{align}
\int_{-1}^{1}dx\,xH^q(x,\xi,t)
&=
A^q(t)+\xi^2D^q(t),
\label{eq:H-second-moment-preview}\\
\int_{-1}^{1}dx\,xE^q(x,\xi,t)
&=
B^q(t)-\xi^2D^q(t).
\label{eq:E-second-moment-preview}
\end{align}
These relations anticipate the energy--momentum-tensor discussion in Sec.~\ref{sec:gravitational}. The individual quark form factors $A^q$, $B^q$, and $D^q$ inherit the renormalization-scale and scheme dependence of the corresponding quark operators; these arguments are suppressed here for notational simplicity. The cancellation of the $D$-term in the sum of Eqs.~\eqref{eq:H-second-moment-preview} and~\eqref{eq:E-second-moment-preview} ensures that the angular-momentum combination is independent of $\xi$.

\subsection{Electromagnetic, Compton, and energy--momentum weightings}
\label{sec:flavor-weightings}

We define the flavor form factors without including the quark electric charges:
\begin{align}
F_1^q(t)
&=
\int_{-1}^{1}dx\,H^q(x,\xi,t),
\label{eq:F1q-sum-rule}\\
F_2^q(t)
&=
\int_{-1}^{1}dx\,E^q(x,\xi,t).
\label{eq:F2q-sum-rule}
\end{align}
Lorentz covariance and current conservation make these first moments independent of $\xi$. Neglecting strange and heavier-flavor contributions and assuming charge symmetry, the proton and neutron form factors are
\begin{align}
F_i^p
&=
\frac{2}{3}F_i^u-\frac{1}{3}F_i^d,
&
F_i^n
&=
\frac{2}{3}F_i^d-\frac{1}{3}F_i^u,
\qquad i=1,2,
\label{eq:nucleon-flavor-form-factors}
\end{align}
and hence
\begin{align}
F_i^u &=2F_i^p+F_i^n,
&
F_i^d &=F_i^p+2F_i^n.
\label{eq:flavor-separated-form-factors}
\end{align}
The latter relations show explicitly why neutron measurements are essential for flavor separation and why correlations between proton and neutron inputs propagate nontrivially into the extracted flavor quantities.

For later convenience, we may also introduce the charge-weighted proton and neutron combinations
\begin{align}
H_{\mathrm{em}}^p
&=
\frac{2H^u-H^d}{3},
&
H_{\mathrm{em}}^n
&=
\frac{2H^d-H^u}{3},
\label{eq:em-H-combinations}
\end{align}
with analogous definitions for $E_{\mathrm{em}}^{p,n}$. These are linear combinations of flavor GPDs, not additional independent distributions. Their first moments reproduce the corresponding electromagnetic form factors.

At this point, it is important to distinguish three physically different flavor combinations. The electromagnetic current weights each quark flavor with its electric charge $e_q$, giving the factors $2/3$ and $-1/3$ for the $u$ and $d$ contributions, respectively. At leading order, the quark contribution to the DVCS amplitude is instead weighted by $e_q^2$, so that the corresponding $u$- and $d$-quark factors are $4/9$ and $1/9$. Finally, the total energy--momentum tensor contains the unit-weighted sum over all quark flavors together with the gluon contribution. These three distinct weightings and the information associated with them are summarized in Table~\ref{tab:flavor-weights}.

\begin{table*}[!t]
\caption{\label{tab:flavor-weights} Flavor weights in the proton $u,d$ sector. At leading order, the quark contribution to DVCS is weighted by $e_q^2$; gluon GPDs enter the hard-scattering amplitude starting at higher orders in $\alpha_s$. The total energy--momentum tensor contains the unit-weighted sum over all quark flavors and the gluon contribution.}
\begin{ruledtabular}
\begin{tabular}{lccc}
Quantity & $u$ weight & $d$ weight & Information accessed\\
\hline
Electromagnetic current
 & $2/3$ & $-1/3$ & $F_{1,2}^{p}$\\
DVCS amplitude at LO
 & $4/9$ & $1/9$ & Compton convolution\\
Total energy--momentum tensor
 & $1$ & $1$ & All quark flavors and gluons
\end{tabular}
\end{ruledtabular}
\end{table*}

The second moments of $H_{\mathrm{em}}^{p,n}$ and $E_{\mathrm{em}}^{p,n}$ are therefore charge-weighted combinations of quark energy--momentum-tensor form factors. They must not be identified with the form factors of the complete proton energy--momentum tensor or with the total proton angular momentum. Those quantities involve unit flavor weights and, at a generic renormalization scale, contributions from all active quark flavors and gluons.

A related distinction concerns conservation. Individual quark-flavor and gluon contributions to the energy--momentum tensor depend on the renormalization scale and are not separately conserved. Conservation applies only to their complete sum. Consequently, the constraints
\begin{equation}
\sum_{a=q,g} A^a(0)=1,
\qquad
\sum_{a=q,g} B^a(0)=0,
\label{eq:full-emt-sum-rules}
\end{equation}
refer to the full energy--momentum tensor, where the quark sum extends over all active flavors and includes the corresponding antiquark contributions. When these relations are imposed within the valence-only low-scale model used below, they constitute explicit matching assumptions about the effective degrees of freedom and do not follow from the electromagnetic sum rules.


\section{A zero-skewness construction}
\label{sec:zero-skewness}

\subsection{Longitudinal input and its scale}
\label{sec:longitudinal-input}

The low-scale interpretation adopted here is motivated by dressed-parton descriptions in which valence degrees of freedom carry most or all of the proton longitudinal momentum at a hadronic matching scale~\cite{Chang2022}. Covariant-extension methods have moreover shown how GPDs constructed from low-Fock-sector overlap representations can be completed while preserving polynomiality~\cite{Chouika2018}. This is a boundary condition for an effective description, not a statement that gluonic interactions or gluonic binding effects are absent. Relating these effective degrees of freedom to renormalized quark and gluon operators at a perturbative scale requires a specified matching and evolution prescription.

For each valence flavor, we use the beta-function form
\begin{equation}
q_v(x) = C_q x^{a_q}(1-x)^{b_q},
\qquad 0<x<1,
\label{eq:valence-input}
\end{equation}
with
\begin{equation}
(a_u,b_u)=(1.15,3.10),
\qquad
(a_d,b_d)=(0.90,3.18),
\label{eq:valence-exponents}
\end{equation}
the model parameters obtained by fitting the PDF results reported in Ref.~\cite{Chang2022}. The coefficients are fixed by exact valence-number normalization:
\begin{equation}
C_q
=
\frac{n_q}{B(a_q+1,b_q+1)},
\qquad
n_u=2,
\qquad
n_d=1,
\label{eq:valence-normalization}
\end{equation}
where $B$ denotes the Euler beta function. All forward moments then follow analytically:
\begin{equation}
\int_0^1 dx\,x^n q_v(x) = C_q B(a_q+n+1,b_q+1).
\label{eq:valence-moments}
\end{equation}
In particular, $n=0$ gives the valence charges, whereas $n=1$ gives the longitudinal momentum fractions carried by the modeled flavors.

Exact quark-number normalization does not automatically imply the momentum condition
\begin{equation}
\sum_q\int_0^1dx\,xq_v(x)=1.
\label{eq:momentum-condition}
\end{equation}
If Eq.~\eqref{eq:momentum-condition} is imposed, it constitutes an additional low-scale matching condition and must be implemented through the longitudinal shape parameters, rather than through a second independent normalization of $C_u$ and $C_d$.

We denote the input scale by $\mu_0$. The nominal value $\mu_0\simeq0.331~\mathrm{GeV}$ should be understood as a hadronic model scale whose meaning depends on the adopted evolution prescription. Since no independent perturbative matching calculation is performed here, all distributions and moments obtained directly from Eqs.~\eqref{eq:valence-input}--\eqref{eq:valence-moments} are identified as low-scale results. Their evolution to conventional factorization scales lies beyond the present analysis.


\subsection{Zero-skewness GPD ansatz}
\label{sec:zero-skewness-ansatz}

At zero skewness, we write the valence GPDs as
\begin{align}
H_v^q(x,t)
&=
q_v(x)\exp\!\left[t f_q(x)\right],
\label{eq:Hv-zero-skewness}\\
E_v^q(x,t)
&=
e_v^q(x)\exp\!\left[t g_q(x)\right].
\label{eq:Ev-zero-skewness}
\end{align}
Although the forward functions and the momentum-transfer dependence appear as separate factors, Eqs.~\eqref{eq:Hv-zero-skewness} and~\eqref{eq:Ev-zero-skewness} do not factorize the $x$ and $t$ dependences because the profile functions $f_q$ and $g_q$ depend on $x$. This correlation permits elastic form factors to constrain the transverse localization of partons carrying different longitudinal momentum fractions.

The forward limit of $E_v^q$ is not determined by an ordinary PDF. We parametrize it independently as
\begin{equation}
 e_v^q(x) = \frac{\kappa_q}{\mathcal N_q}\, x^{\alpha_q}(1-x)^{\beta_q} \left(1+\gamma_q x\right),
 \label{eq:forward-E-flexible}
\end{equation}
where
\begin{equation}
\mathcal N_q = B(\alpha_q+1,\beta_q+1) + \gamma_q B(\alpha_q+2,\beta_q+1).
\label{eq:E-normalization}
\end{equation}
It follows identically that
\begin{equation}
\int_0^1dx\,e_v^q(x)=\kappa_q,
\label{eq:kappa-normalization}
\end{equation}
with
\begin{equation}
\kappa_u=1.673, \qquad \kappa_d=-2.033.
\end{equation}
Thus the anomalous-magnetic-moment normalization is separated from the shape of $e_v^q$.

The first $x$-weighted moment is
\begin{equation}
B^q(0) = \frac{\kappa_q}{\mathcal N_q} \left[ B(\alpha_q+2,\beta_q+1) + \gamma_q B(\alpha_q+3,\beta_q+1) \right].
\label{eq:Bq-zero}
\end{equation}
Within the quark-only low-scale description, one combination of the Pauli-shape parameters may be eliminated by imposing
\begin{equation}
B^u(0)+B^d(0)=0.
\label{eq:B-total-zero}
\end{equation}
Equation~\eqref{eq:B-total-zero} is the vanishing anomalous gravitomagnetic-moment condition in the adopted effective description. It should not itself be called Ji's sum rule. Ji's relation additionally involves $A^q(0)$:
\begin{equation}
J^q = \frac{1}{2}\left[A^q(0)+B^q(0)\right].
\label{eq:Ji-reminder}
\end{equation}
Consequently, $B^{u+d}(0)=0$ implies $J^{u+d}=A^{u+d}(0)/2$. Saturation of the proton spin by the total angular momentum of the modeled quark degrees of freedom follows only if $A^{u+d}(0)=1$ is imposed independently. No separate decomposition into quark-spin and quark-orbital contributions is obtained from the unpolarized GPDs alone.

For the transverse profiles, we use
\begin{align}
f_q(x)
 &=
 \alpha'_{H,q}(1-x)^3\ln(1/x)
 +
 a_{H,q}\,x(1-x)^2 \nonumber \\
 &+
 b_{H,q}(1-x)^3,
 \label{eq:H-profile}\\
 g_q(x)
 &=
 \alpha'_{E,q}(1-x)^3\ln(1/x)
 +
 a_{E,q}\,x(1-x)^2 \nonumber \\
 &+
 b_{E,q}(1-x)^3.
 \label{eq:E-profile}
\end{align}
All coefficients in Eqs.~\eqref{eq:H-profile} and~\eqref{eq:E-profile} have dimensions of \(\mathrm{GeV}^{-2}\). This notation avoids confusing the profile parameters with the energy--momentum-tensor form factors $A^q(t)$ and $B^q(t)$.

At small $x$,
\begin{align}
f_q(x) &\sim \alpha'_{H,q}\ln(1/x)+b_{H,q},\\
g_q(x) &\sim \alpha'_{E,q}\ln(1/x)+b_{E,q},
\end{align}
whereas at large $x$,
\begin{align}
f_q(x)&\sim a_{H,q}(1-x)^2,\\
g_q(x)&\sim a_{E,q}(1-x)^2,
\end{align}
provided the corresponding $a$ coefficients are nonzero. The logarithmic contribution is motivated by Regge-like behavior, but a fitted low-scale coefficient should not be interpreted directly as a measured Regge slope. The numerical profile parameters are determined in Sec.~\ref{sec:electromagnetic} from the simultaneous electromagnetic calibration.


\subsection{Endpoint behavior, positivity, and large momentum transfer}
\label{sec:endpoint-behavior}

The endpoint behavior provides a useful analytic connection between the large-$x$ longitudinal distribution and the large-$Q^2$ behavior of the elastic form factors. Let $y=1-x$, and suppose
\begin{equation}
q_v(x)\sim C_q y^{b_q},
\qquad
f_q(x)\sim a_{H,q}y^2,
\end{equation}
as $y\to0$. With $Q^2=-t>0$, the leading endpoint contribution is
\begin{align}
F_1^q(-Q^2) & \sim C_q\int_0^\infty dy\,y^{b_q} \exp\!\left[-a_{H,q}Q^2y^2\right]
\nonumber\\
&=
\frac{C_q}{2} \Gamma\!\left(\frac{b_q+1}{2}\right) \left(a_{H,q}Q^2\right)^{-(b_q+1)/2}.
\label{eq:F1-endpoint}
\end{align}
For $b_q\simeq3$, this behaves approximately as $1/Q^4$.

Similarly, if
\begin{equation}
e_v^q(x)\sim \widetilde C_q y^{\beta_q},
\qquad
g_q(x)\sim a_{E,q}y^2,
\end{equation}
with $\widetilde C_q = \kappa_q(1+\gamma_q)/{\cal N}_q$, then
\begin{equation}
 F_2^q(-Q^2)
 \sim
 \frac{\widetilde C_q}{2}
 \Gamma\!\left(\frac{\beta_q+1}{2}\right)
 \left(a_{E,q}Q^2\right)^{-(\beta_q+1)/2}.
 \label{eq:F2-endpoint}
\end{equation}
The form-factor ratio therefore behaves as
\begin{equation}
\frac{F_2^q(-Q^2)}{F_1^q(-Q^2)}
\sim
\left(Q^2\right)^{-(\beta_q-b_q)/2},
\label{eq:F2-F1-endpoint}
\end{equation}
up to flavor-dependent coefficients. A proportional choice $e_v^q\propto q_v$ would give $\beta_q=b_q$ and would not produce any additional power suppression of $F_2^q/F_1^q$. The often-used additional $1/Q^2$ suppression corresponds, within this Gaussian endpoint analysis, to $\beta_q-b_q=2$~\cite{Diehl2005, DiehlKroll2013}. Agreement over a finite fitted interval should not be interpreted as proof that the true asymptotic regime has been reached.

The endpoint powers are also constrained by positivity. For the Gaussian profiles used here, the impact-parameter inequality derived in Sec.~\ref{sec:impact} implies, when $g_q/f_q$ tends to a finite value strictly between zero and one,
\begin{equation}
\frac{|e_v^q(x)|}{q_v(x)} = \mathcal O(1-x) \quad\text{or faster} \qquad (x\to1).
\label{eq:E-H-endpoint-positivity}
\end{equation}
The parametrizations in Eqs.~\eqref{eq:valence-input} and
\eqref{eq:forward-E-flexible} therefore require
\begin{equation}
\beta_q\geq b_q+1.
\label{eq:beta-positivity}
\end{equation}
This condition is necessary but not sufficient for pointwise positivity over the complete interval $0<x<1$. The full inequality must therefore be imposed
or verified during the constrained calibration. In particular, the profile ordering $0<g_q(x)<f_q(x)$ alone does not guarantee positivity of the transversely polarized density.


\section{Electromagnetic calibration and parameter determination}
\label{sec:electromagnetic}

\subsection{Electromagnetic observables and calibration strategy}
\label{sec:em-calibration-strategy}

At zero skewness, the flavor-separated Dirac and Pauli form factors follow from the GPD sum rules,
\begin{align}
F_1^q(t) &= \int_0^1\dd x\,q_v(x)\exp\!\left[t f_q(x)\right],
\label{eq:F1q-zero-skewness}
\\
F_2^q(t) &= \int_0^1\dd x\,e_v^q(x)\exp\!\left[t g_q(x)\right].
\label{eq:F2q-zero-skewness}
\end{align}
Under charge symmetry and neglecting strange-current contributions, the proton and neutron form factors used in the calibration are constructed from the flavor-separated quantities according to Eq.~\eqref{eq:nucleon-flavor-form-factors}. The corresponding Sachs form factors are
\begin{align}
G_E^N(t) &= F_1^N(t)+\frac{t}{4M^2}F_2^N(t), \label{eq:Sachs-electric}\\ 
G_M^N(t) &= F_1^N(t)+F_2^N(t),
\label{eq:Sachs-magnetic}
\end{align}
where $N=p,n$. We use the common nucleon mass $M=0.938272~\mathrm{GeV}$, consistently with the charge-symmetric approximation.

The purpose of the electromagnetic calibration is to constrain the parameters entering the transverse profiles $f_q$ and $g_q$, together with the shape of the forward helicity-flip distribution $e_v^q$. The longitudinal distributions $q_v$ and their exponents are fixed as described in Sec.~\ref{sec:longitudinal-input}. Before implementing the exact constraints discussed below, the complete parameter set is
\begin{equation}
\boldsymbol{\theta}
=
\left\{
\alpha'_{H,q},a_{H,q},b_{H,q},
\alpha'_{E,q},a_{E,q},b_{E,q},
\alpha_q,\beta_q,\gamma_q
\right\}_{q=u,d}.
\label{eq:fit-parameter-vector}
\end{equation}
The parameter $\gamma_d$ is eliminated by the condition $B^u(0)+B^d(0)=0$. The remaining parameters are varied simultaneously subject to the endpoint and positivity constraints. Thus, $17$ independent parameters are optimized in the final calibration.

We use the Kelly parametrization~\cite{Kelly2004} as a common phenomenological reference for the four nucleon electromagnetic form factors. The reference Dirac and Pauli form factors are constructed from the corresponding Sachs functions according to
\begin{align}
F_{1,\mathrm{K}}^N(Q^2)
&=
\frac{G_{E,\mathrm{K}}^N(Q^2) +\tau G_{M,\mathrm{K}}^N(Q^2)} {1+\tau},
\label{eq:Kelly-F1}\\
F_{2,\mathrm{K}}^N(Q^2) &= \frac{G_{M,\mathrm{K}}^N(Q^2) -G_{E,\mathrm{K}}^N(Q^2)} {1+\tau},
\label{eq:Kelly-F2}\\
\tau&=\frac{Q^2}{4M^2}.
\nonumber
\end{align}
The calibration is performed over
\begin{equation}
0.01\leq Q^2\leq4.0~\mathrm{GeV}^2,
\qquad Q^2=-t.
\label{eq:fit-Q2-range}
\end{equation}
Low-$Q^2$ anchor points are included in order to retain sensitivity to the slopes at the origin.

It is important to distinguish this procedure from a direct fit to experimental measurements. Values sampled from a single analytic parametrization are correlated predictions of a common set of parameters and must not be treated as statistically independent measurements. Since the full covariance matrix of the reference parametrization is not used here, the minimized quantity is a weighted calibration functional rather than a statistical chi-squared:
\begin{align}
\Phi_{\mathrm{em}}(\boldsymbol{\theta}) &= \sum_{N=p,n} \sum_{i=1,2} \sum_k w_{Ni,k} \nonumber \\
&
\times \left[ F_{i,\mathrm{model}}^N(Q_k^2;\boldsymbol{\theta}) - F_{i,\mathrm{K}}^N(Q_k^2) \right]^2.
\label{eq:calibration-functional}
\end{align}
The weights \(w_{Ni,k}\) define the relative importance assigned to each nucleon, form-factor channel, and momentum-transfer point. In the present calibration, the proton and neutron reference curves are assigned nominal relative tolerances of $5\%$ and $10\%$, respectively. The larger neutron tolerance reflects the enhanced sensitivity of neutron form factors to cancellations between the $u$- and $d$-quark contributions. For channels that vanish at the origin or cross zero, the numerical weights are regularized with a nonzero channel scale so that the calibration functional does not develop a spurious singularity.

The low-$Q^2$ behavior is further guided by a proton electric-radius term,
\begin{equation}
\Phi(\boldsymbol{\theta})
=
\Phi_{\mathrm{em}}(\boldsymbol{\theta})
+
\lambda_E \left[ \frac{r_E^p(\boldsymbol{\theta})-r_{E,\mathrm{ref}}^p}{1~\mathrm{fm}} \right]^2,
\quad
r_{E,\mathrm{ref}}^p=0.84~\mathrm{fm},
\label{eq:total-calibration-functional}
\end{equation}
where $\lambda_E$ specifies the numerical strength of the radius penalty. This term guides the calibration toward the empirical proton-radius region but does not override the exact GPD constraints. The resulting functional $\Phi$ is an optimization criterion; it must not be interpreted as a statistical chi-squared, and variations derived from it are not experimental confidence intervals.


\subsection{Exact constraints and positivity conditions}
\label{sec:fit-constraints}

Several constraints are implemented before or during the electromagnetic minimization. The valence-number conditions
\begin{equation}
\int_0^1dx\,u_v(x)=2,
\qquad
\int_0^1dx\,d_v(x)=1
\label{eq:fit-valence-normalization}
\end{equation}
are built into the coefficients $C_u$ and $C_d$. Similarly, the normalization of the forward helicity-flip distribution guarantees
\begin{equation}
 \int_0^1dx\,e_v^q(x)=\kappa_q,
\label{eq:fit-Pauli-normalization}
\end{equation}
and hence reproduces the proton and neutron anomalous magnetic moments at $t=0$.

Within the adopted valence-only low-scale description, the parameter $\gamma_d$ is determined by imposing
\begin{equation}
B^u(0)+B^d(0)=0,
\label{eq:fit-B-zero}
\end{equation}
where $B^q(0)$ is given by Eq.~\eqref{eq:Bq-zero}. This is an independent $x$-weighted condition and does not follow from the Pauli normalization in Eq.~\eqref{eq:fit-Pauli-normalization}. It implies
\begin{equation}
J^{u+d}
=
\frac{1}{2}A^{u+d}(0),
\end{equation}
but it does not by itself imply $J^{u+d}=1/2$. Saturation of the proton spin by the total angular momentum of the modeled quark degrees of freedom additionally requires the low-scale momentum condition $A^{u+d}(0)=1$.

The endpoint requirement derived in
Sec.~\ref{sec:endpoint-behavior} is imposed through
\begin{equation}
\beta_q\geq b_q+1.
\label{eq:fit-beta-bound}
\end{equation}
The transverse profiles are required to obey
\begin{equation}
f_q(x)>0, \qquad 0<g_q(x)<f_q(x), \qquad 0<x<1.
\label{eq:profile-ordering}
\end{equation}
These conditions are necessary but not sufficient for positivity of the transversely polarized density. Accordingly, we also impose the complete Gaussian inequality derived in Sec.~\ref{sec:impact},
\begin{equation}
[e_v^q(x)]^2 \leq 8eM^2 \left[\frac{g_q(x)}{f_q(x)}\right]^3 \left[f_q(x)-g_q(x)\right] [q_v(x)]^2,
\label{eq:fit-full-positivity}
\end{equation}
for both flavors over the complete interval $0<x<1$.

In the numerical implementation, Eq.~\eqref{eq:fit-full-positivity} is evaluated on a grid refined near both endpoints and supplemented by the analytic condition in Eq.~\eqref{eq:fit-beta-bound}. The optimization accepts a parameter set only when the valence and Pauli normalizations, the condition $B^{u+d}(0)=0$, the endpoint bound, the profile ordering, and the full Gaussian positivity inequality are simultaneously satisfied. Thus, the radius penalty and electromagnetic residuals select among theoretically admissible parametrizations; they are not allowed to compensate for a violation of the exact GPD constraints.

\begin{table*}[!t]
\caption{\label{tab:optimized-parameters} Parameters obtained from the simultaneous constrained electromagnetic calibration. Profile parameters have units of $\mathrm{GeV}^{-2}$; the exponents and $\gamma_q$ are dimensionless. The parameter $\gamma_d$, marked ``derived,'' is fixed by $B^u(0)+B^d(0)=0$. All remaining parameters are varied subject to the endpoint, profile-ordering, and Gaussian positivity constraints.}
\begin{ruledtabular}
\begin{tabular}{cccccccccc}
Flavor &
$\alpha'_{H,q}$ & $a_{H,q}$ & $b_{H,q}$ & $\alpha'_{E,q}$ & $a_{E,q}$ & $b_{E,q}$ & $\alpha_q$ & $\beta_q$ & $\gamma_q$ \\
\hline
$u$ & 1.983 & 2.856 & 2.320 & 1.888 & 2.505 & 2.439 & 1.186 & 4.694 & 3.074\\
$d$ & 1.918 & 3.993 & 2.371 & 1.708 & 2.419 & 1.944 & 0.909 & 4.710 & 0.434 \text{ (derived)}
\end{tabular}
\end{ruledtabular}
\end{table*}

\begin{figure*}[t]
\includegraphics[width=.90\textwidth]{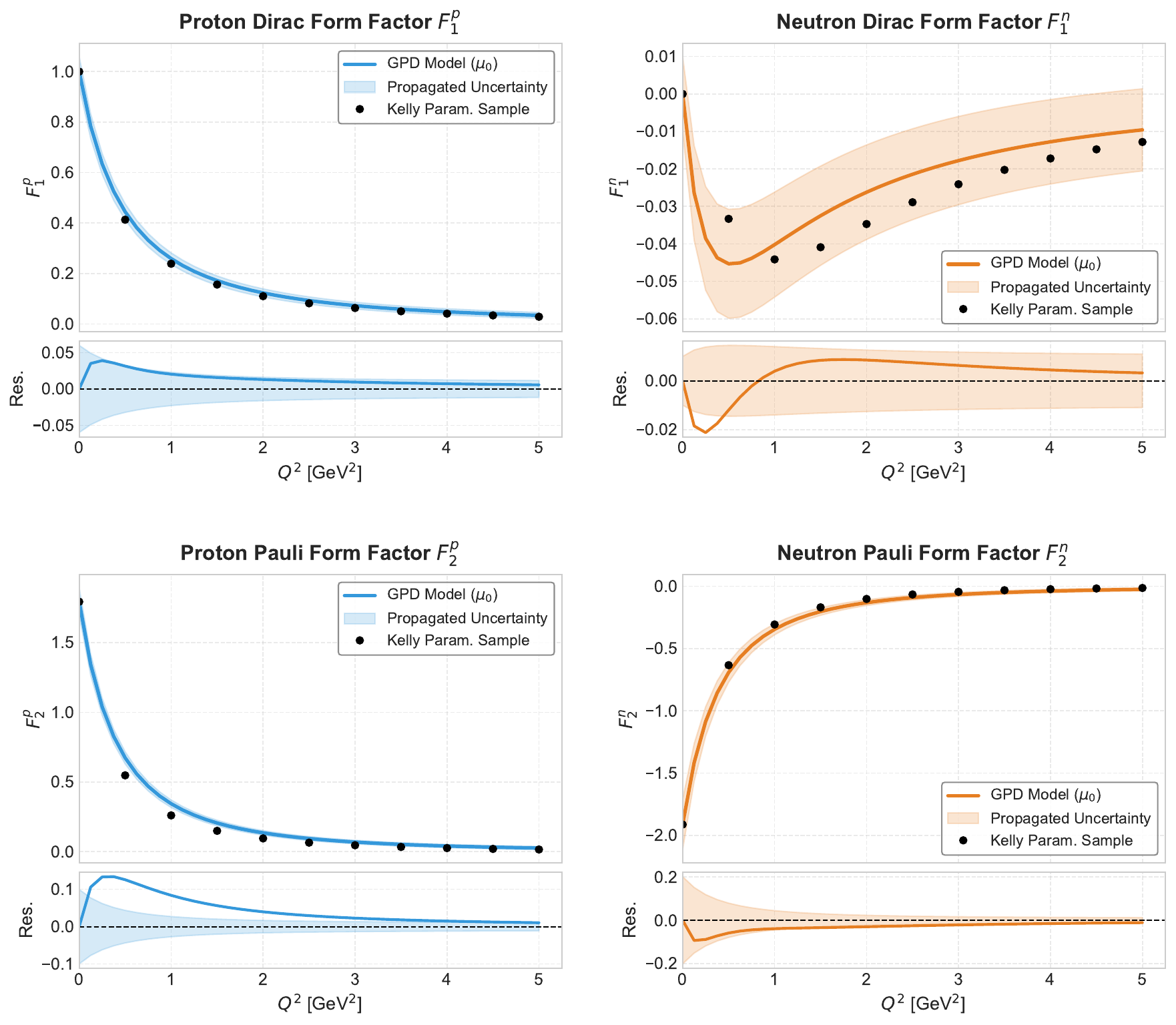}
\caption{\label{fig:electromagnetic-FFs} Proton and neutron Dirac and Pauli form factors obtained from the simultaneous constrained electromagnetic calibration. Solid curves show the reconstructed form factors at the hadronic input scale $\mu_0$, and the markers are values sampled from the central Kelly parametrization; the latter are not independent experimental measurements. The calibration is performed over $0.01\leq Q^2\leq4.0~\mathrm{GeV}^2$; the continuation to $Q^2=5.0~\mathrm{GeV}^2$ displays the behavior immediately outside the calibrated interval. The lower panels show the absolute residuals, $F_{i,\mathrm{model}}^N-F_{i,\mathrm{K}}^N$. Absolute rather than percentage residuals are displayed because neutron form factor vanishes at the origin or remains small as a result of flavor cancellations; dividing by the reference form factor would therefore produce singular or visually misleading relative deviations. The shaded regions represent sensitivity to the adopted calibration tolerances within the fitted ansatz and should not be interpreted as experimental confidence intervals.}
\end{figure*}

\subsection{Optimized parameters and form-factor description}
\label{sec:optimized-em-parameters}

The simultaneous constrained calibration produces the parameter set listed in
Table~\ref{tab:optimized-parameters}. The resulting proton and neutron Dirac and Pauli form factors are compared with the Kelly reference curves in Fig.~\ref{fig:electromagnetic-FFs}. Because the proton and neutron channels are calibrated simultaneously, this comparison tests both the overall electromagnetic description and the flavor separation implied by charge symmetry. The neutron form factors are particularly restrictive because they result from partial cancellations between the $u$- and $d$-quark contributions.

The constrained parametrization reproduces the principal momentum-transfer dependence of the four form factors, but visible residuals remain, especially at low and intermediate $Q^2$. These deviations are not hidden by retuning parameters outside the theoretically admissible domain. They quantify the compromise enforced by the restricted functional ansatz between the electromagnetic reference curves, the low-$Q^2$ slopes, the endpoint behavior, and impact-parameter positivity. The result should therefore be interpreted as a theoretically consistent GPD reconstruction with a reasonable electromagnetic description, rather than as a precision global fit of nucleon form-factor data.

The shaded regions in Fig.~\ref{fig:electromagnetic-FFs} display the parameter sensitivity obtained by propagating the adopted calibration tolerances through the fitted ansatz. Since the calibration uses phenomenological central curves and user-defined channel weights rather than the experimental covariance matrix, these regions describe the local sensitivity of the reconstruction to the adopted calibration procedure. They are not statistical confidence intervals and do not include the full experimental, parametrization, PDF, matching-scale, or functional-form uncertainties.


\subsection{Electromagnetic slopes and radii}
\label{sec:em-radii}

Differentiating Eqs.~\eqref{eq:F1q-zero-skewness} and~\eqref{eq:F2q-zero-skewness} at $t=0$ gives
\begin{align}
F_1^{q\,\prime}(0) &= \int_0^1dx\,q_v(x)f_q(x), \label{eq:F1-slope}\\
F_2^{q\,\prime}(0) &= \int_0^1dx\,e_v^q(x)g_q(x).
\label{eq:F2-slope}
\end{align}
The slope of the electric Sachs form factor contains both Dirac and Pauli contributions:
\begin{equation}
G_E^{N\,\prime}(0) = F_1^{N\,\prime}(0) + \frac{F_2^N(0)}{4M^2}.
\label{eq:GE-slope}
\end{equation}
The second term in Eq.~\eqref{eq:GE-slope} must be retained for both the proton and neutron.

In the present $t$-convention, the electromagnetic radii are defined by
\begin{align}
(r_E^p)^2 &= 6G_E^{p\,\prime}(0),
\label{eq:proton-electric-radius}\\
\langle r_E^2\rangle_n &= 6G_E^{n\,\prime}(0),
\label{eq:neutron-charge-radius}\\
(r_M^N)^2 &= \frac{6}{G_M^N(0)} G_M^{N\,\prime}(0).
\label{eq:magnetic-radius}
\end{align}
Since $G_E^n(0)=0$, the neutron mean-square charge radius is defined directly from the slope, without division by the value of the form factor at the origin; it may therefore be negative.

For the positivity-compatible solution in Table~\ref{tab:optimized-parameters}, we obtain
\begin{align}
r_E^p &= 0.78122~\mathrm{fm}, & \langle r_E^2\rangle_n &= -0.05625~\mathrm{fm}^2, \nonumber\\
r_M^p &= 0.74052~\mathrm{fm}, & r_M^n &= 0.75168~\mathrm{fm}.
\label{eq:new-radii}
\end{align}
These radii are derived consequences of the constrained calibration and its behavior near $t=0$. They are not independent predictions because the calibration includes low-$Q^2$ anchor points and an explicit proton-radius penalty. Conversely, they are not direct fits to radius measurements, and no statistical uncertainties are assigned to them.

Despite the term favoring $r_E^p\simeq0.84~\mathrm{fm}$, the constrained optimum produces the smaller value shown in Eq.~\eqref{eq:new-radii}. Increasing the radius further within the adopted functional form generates tension with the finite-$Q^2$ form factors or with the complete Gaussian positivity bound. The radii therefore provide a particularly transparent diagnostic of the limitations of the ansatz: the final solution prioritizes exact normalization, endpoint behavior, angular-momentum matching, and impact-parameter positivity over a precision description of every low-$Q^2$ electromagnetic observable.


\section{Transverse localization and positivity}
\label{sec:impact}

\subsection{Impact-parameter representation}
\label{sec:impact-parameter-representation}

At zero skewness, the transverse Fourier transform of the quark GPD $H^q(x,0,t)$ for $x>0$ admits a density interpretation in a proton state localized in the transverse plane~\cite{Burkardt2000, Diehl2002}. In the low-scale valence-only description adopted in this work, the antiquark contribution is neglected and $H^q$ is represented by the valence distribution $H_v^q$. Within this approximation, we define
\begin{equation}
q_v(x,\boldsymbol b_\perp)
=
\int\frac{d^2\boldsymbol\Delta_\perp}{(2\pi)^2}\, e^{-i\boldsymbol b_\perp\cdot\boldsymbol\Delta_\perp} H_v^q(x,0,-\boldsymbol\Delta_\perp^2),
\label{eq:impact-H-definition}
\end{equation}
where $\boldsymbol b_\perp$ is measured from the light-front transverse center of momentum. For the exponential zero-skewness ansatz of Eq.~\eqref{eq:Hv-zero-skewness}, the transform is Gaussian:
\begin{equation}
q_v(x,\boldsymbol b_\perp)
=
\frac{q_v(x)}{4\pi f_q(x)}
\exp\left[-\frac{\boldsymbol b_\perp^2}{4f_q(x)}\right].
\label{eq:impact-H-Gaussian}
\end{equation}
Provided that $q_v(x)\geq0$ and $f_q(x)>0$, this distribution is
nonnegative and satisfies
\begin{equation}
\int d^2\boldsymbol b_\perp\,
q_v(x,\boldsymbol b_\perp)
=
q_v(x).
\label{eq:impact-H-normalization}
\end{equation}
Its conditional transverse mean-square position is
\begin{equation}
\left\langle b_\perp^2\right\rangle_x
=
\frac{\int d^2\boldsymbol b_\perp\,\boldsymbol b_\perp^2 q_v(x,\boldsymbol b_\perp)}{\int d^2\boldsymbol b_\perp\, q_v(x,\boldsymbol b_\perp)} =
4f_q(x).
\label{eq:conditional-impact-radius}
\end{equation}

Equation~\eqref{eq:conditional-impact-radius} describes transverse localization in a light-front state. It is not a three-dimensional rest-frame radius and should not be identified with a Sachs radius. In particular, replacing $b_\perp$ by a three-dimensional radial coordinate would change the physical meaning of the distribution.

The transverse position of the active quark relative to the transverse center of momentum of the spectators scales as $\boldsymbol b_\perp/(1-x)$. Hence,
\begin{equation}
\left\langle \frac{b_\perp^2}{(1-x)^2} \right\rangle_x = \frac{4f_q(x)}{(1-x)^2}.
\label{eq:active-spectator-separation}
\end{equation}
Since the fitted profiles behave as $f_q(x)\sim a_{H,q}(1-x)^2$ when $x\to1$, the right-hand side approaches a finite value. Thus, as $x\to1$, the active quark approaches the transverse center of momentum while its separation from the transverse center of momentum of the spectators remains finite.

For the helicity-flip GPD, we analogously define
\begin{equation}
\mathcal E_v^q(x,\boldsymbol b_\perp) = \int\frac{d^2\boldsymbol\Delta_\perp}{(2\pi)^2}\, e^{-i\boldsymbol b_\perp\cdot\boldsymbol\Delta_\perp} E_v^q(x,0,-\boldsymbol\Delta_\perp^2),
\label{eq:impact-E-definition}
\end{equation}
which gives
\begin{equation}
\mathcal E_v^q(x,\boldsymbol b_\perp) = \frac{e_v^q(x)}{4\pi g_q(x)} \exp\left[-\frac{\boldsymbol b_\perp^2}{4g_q(x)}\right].
\label{eq:impact-E-Gaussian}
\end{equation}
Unlike Eq.~\eqref{eq:impact-H-Gaussian}, $\mathcal E_v^q$ is not itself a probability density. Its transverse derivative describes the sideways distortion of the unpolarized valence-quark distribution in a transversely polarized proton.

For a proton polarized along $+\hat{\boldsymbol x}$, our Fourier and spin conventions give
\begin{equation}
q_{v,+\hat x}^q(x,\boldsymbol b_\perp) = q_v(x,\boldsymbol b_\perp)
- \frac{1}{2M} \frac{\partial}{\partial b_y} \mathcal E_v^q(x,\boldsymbol b_\perp).
\label{eq:transversely-polarized-density}
\end{equation}
Reversing the transverse polarization reverses the sign of the derivative term. At fixed $x$, the corresponding average sideways shift is
\begin{equation}
\left\langle b_y\right\rangle_x^q = \frac{
\int d^2\boldsymbol b_\perp\, b_y q_{v,+\hat x}^q(x,\boldsymbol b_\perp)}{\int d^2\boldsymbol b_\perp\, q_{v,+\hat x}^q(x,\boldsymbol b_\perp)} = \frac{e_v^q(x)}{2M q_v(x)}.
\label{eq:fixed-x-transverse-shift}
\end{equation}
Thus, within the stated Fourier and polarization conventions, the sign of $e_v^q$ fixes the direction of the flavor-dependent distortion. After integration over $x$, one obtains
\begin{equation}
\left\langle b_y\right\rangle^q = \frac{\kappa_q}{2Mn_q}, \qquad n_u=2, \qquad n_d=1.
\label{eq:integrated-transverse-shift}
\end{equation}
This relation provides a useful normalization check on the numerical
Fourier transforms.


\begin{figure*}[!t]
\includegraphics[width=0.95\textwidth]{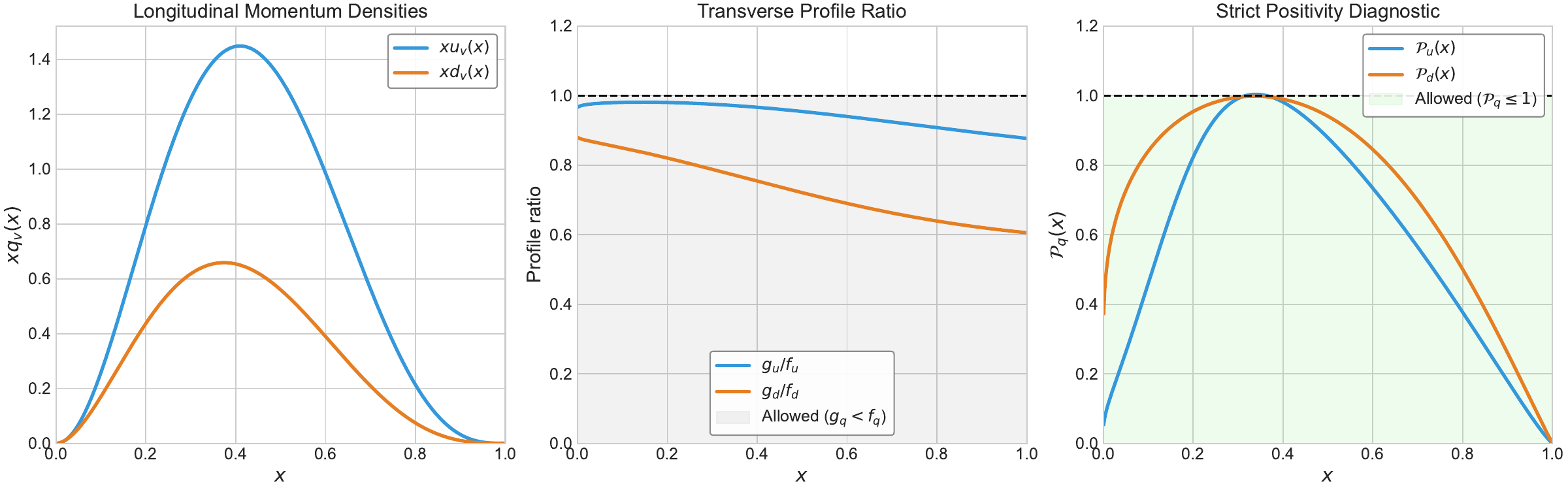}
\caption{\label{fig:longitudinal-and-positivity} Proton longitudinal momentum densities and positivity diagnostics obtained from the constrained electromagnetic calibration. The left panel shows the proton valence distributions $x u_v(x)$ and $x d_v(x)$. The middle panel displays the corresponding profile ratios $g_q(x)/f_q(x)$, with the horizontal dashed line marking the boundary $g_q/f_q=1$. The right panel shows the saturation functions $\mathcal P_q(x)$ defined in Eq.~\eqref{eq:positivity-saturation}, with the horizontal dashed line marking $\mathcal P_q=1$. Nonnegativity of the modeled transversely polarized proton valence density requires both $0<g_q/f_q<1$ and $\mathcal P_q\leq1$.}
\end{figure*}

\subsection{Gaussian positivity criterion}
\label{sec:Gaussian-positivity}

Within the low-scale valence-only interpretation specified above, non-negativity of the unpolarized valence density for either orientation of the transverse proton polarization requires
\begin{equation}
\frac{\left|\boldsymbol\nabla_{\!b} \mathcal E_v^q(x,\boldsymbol b_\perp)\right|
}{2M} \leq q_v(x,\boldsymbol b_\perp).
\label{eq:impact-positivity}
\end{equation}
This condition concerns the particular transverse density constructed from $H_v^q$ and $E_v^q$. More general GPD positivity relations may also involve helicity-dependent distributions, and the interpretation of renormalized parton densities beyond leading order requires additional care~\cite{Pobylitsa2002}.

For the Gaussian functions in Eqs.~\eqref{eq:impact-H-Gaussian} and \eqref{eq:impact-E-Gaussian}, the ratio of the two sides of Eq.~\eqref{eq:impact-positivity} is
\begin{equation}
R_q(x,b_\perp)
=
\frac{|e_v^q(x)|}{q_v(x)} \frac{b_\perp f_q(x)}{4M g_q^2(x)}
\exp\left\{ \frac{b_\perp^2}{4} \left[ \frac{1}{f_q(x)}-\frac{1}{g_q(x)} \right] \right\}.
\label{eq:Gaussian-positivity-ratio}
\end{equation}
If $g_q(x)>f_q(x)$ and $e_v^q(x)\neq0$, this ratio diverges at large $b_\perp$. For $g_q(x)=f_q(x)$, it grows linearly with $b_\perp$. A necessary condition is therefore
\begin{equation}
0<g_q(x)<f_q(x)
\label{eq:profile-ordering-positivity}
\end{equation}
wherever $e_v^q(x)\neq0$.

When Eq.~\eqref{eq:profile-ordering-positivity} holds,
$R_q(x,b_\perp)$ reaches its maximum at
\begin{equation}
b_{\perp,\mathrm{max}}^2(x)
=
\frac{2f_q(x)g_q(x)}{f_q(x)-g_q(x)}.
\label{eq:bmax-positivity}
\end{equation}
Requiring this maximum not to exceed unity gives
\begin{equation}
[e_v^q(x)]^2
\leq
8eM^2
\left[\frac{g_q(x)}{f_q(x)}\right]^3
\left[f_q(x)-g_q(x)\right]
[q_v(x)]^2,
\label{eq:complete-Gaussian-bound}
\end{equation}
where $e$ is Euler's number. Together with Eq.~\eqref{eq:profile-ordering-positivity},
Eq.~\eqref{eq:complete-Gaussian-bound} is the necessary and sufficient pointwise condition for nonnegativity of the transversely polarized valence density within the present Gaussian ansatz. It should not be identified with the complete set of general GPD positivity bounds, which can additionally involve helicity-dependent distributions.

For numerical validation, it is convenient to define the dimensionless saturation function
\begin{equation}
\mathcal P_q(x) = \frac{
[e_v^q(x)]^2}{8eM^2 \left[g_q(x)/f_q(x)\right]^3 \left[f_q(x)-g_q(x)\right] [q_v(x)]^2}.
\label{eq:positivity-saturation}
\end{equation}
The required condition is
\begin{equation}
0\leq\mathcal P_q(x)\leq1,
\qquad
0<x<1.
\label{eq:positivity-saturation-condition}
\end{equation}
The constrained electromagnetic calibration imposes both Eqs.~\eqref{eq:profile-ordering-positivity} and \eqref{eq:positivity-saturation-condition} for the $u$- and $d$-quark distributions.

For the final parameter set, the numerical maxima evaluated with the
full-precision fit parameters are
\begin{align}
\max_{0<x<1}\mathcal P_u(x) &= 1.00008,
\nonumber\\
\max_{0<x<1}\mathcal P_d(x) &= 0.99853.
\label{eq:max-positivity-values}
\end{align}
Both values must lie below unity. The very small excess above unity in the $u$-quark channel, approximately $8\times10^{-5}$, is at the level of the numerical optimization tolerance. Within that tolerance, the solution saturates the positivity boundary. The $d$-quark distribution remains below the boundary throughout the interval.

Apart from the numerical tolerance discussed above, these results show that the fitted distributions reach, but do not significantly cross, the positivity boundary. Figure~\ref{fig:longitudinal-and-positivity} displays the longitudinal momentum densities together with the two complementary diagnostics: the necessary profile ordering $g_q/f_q<1$ and the complete Gaussian saturation function $\mathcal P_q(x)$.


\subsection{Endpoint consistency}
\label{sec:impact-endpoint-consistency}

The positivity inequality constrains both endpoints of the longitudinal distributions. At small $x$, the parametrizations behave as
\begin{equation}
q_v(x)\sim x^{a_q}, \qquad e_v^q(x)\sim x^{\alpha_q}.
\label{eq:small-x-forward-behavior}
\end{equation}
If the leading logarithmic terms satisfy
$\alpha'_{H,q}>\alpha'_{E,q}>0$, one has
\begin{equation}
f_q(x)-g_q(x) \sim \left(\alpha'_{H,q}-\alpha'_{E,q}\right)\ln(1/x),
\qquad x\to0,
\end{equation}
whereas $g_q(x)/f_q(x)$ approaches a finite positive constant smaller than unity. Equation~\eqref{eq:positivity-saturation} then behaves,
up to an $x$-independent coefficient, as
\begin{equation}
\mathcal P_q(x) \sim \frac{x^{2(\alpha_q-a_q)}}{\ln(1/x)}, \qquad x\to0.
\label{eq:small-x-positivity-behavior}
\end{equation}
Finiteness of the saturation function therefore requires
\begin{equation}
\alpha_q\geq a_q.
\label{eq:impact-alpha-condition}
\end{equation}
If the leading logarithmic coefficients of $f_q$ and $g_q$ coincide, the first nonvanishing term in $f_q-g_q$ must instead be examined explicitly. In all cases, the small-$x$ behavior should be checked analytically in addition to the numerical grid test.

At the opposite endpoint, suppose that
\begin{equation}
f_q(x)\sim a_{H,q}(1-x)^2,
\qquad
g_q(x)\sim a_{E,q}(1-x)^2,
\end{equation}
and that $g_q/f_q$ approaches a finite value strictly between zero and one as $x\to1$. Dividing Eq.~\eqref{eq:complete-Gaussian-bound} by $q_v^2(x)$ then shows that
\begin{equation}
\frac{|e_v^q(x)|}{q_v(x)} = \mathcal O(1-x) \quad\text{or faster}.
\label{eq:impact-endpoint-condition}
\end{equation}
For
\begin{equation}
q_v(x)\sim(1-x)^{b_q}, \qquad e_v^q(x)\sim(1-x)^{\beta_q},
\end{equation}
this requires
\begin{equation}
\beta_q\geq b_q+1.
\label{eq:impact-beta-condition}
\end{equation}

The small- and large-$x$ conditions can therefore be summarized as
\begin{equation}
\alpha_q\geq a_q, \qquad \beta_q\geq b_q+1.
\label{eq:impact-endpoint-bounds}
\end{equation}
These analytic endpoint conditions are necessary but do not replace the verification of the complete inequality over the interior of the interval.

A strictly proportional forward ansatz, $e_v^q(x)\propto q_v(x)$, is incompatible with the large-$x$ Gaussian positivity requirement when the profile ratio approaches a finite value strictly between zero and one. The independently parametrized Pauli distribution introduced in Sec.~\ref{sec:zero-skewness-ansatz} is therefore needed both to satisfy the endpoint constraints and to provide sufficient freedom for the $x$-weighted moment $B^q(0)$.


\section{Energy--momentum-tensor form factors}
\label{sec:gravitational}

\subsection{Operator convention and second Mellin moments}
\label{sec:emt-convention}

Let $T_a^{\mu\nu}$ denote the symmetric Belinfante energy--momentum tensor associated with a quark flavor or with the gluon contribution, collectively labeled by $a$. For a spin-one-half target, we use the decomposition
\cite{Ji1997, PolyakovSchweitzer2018}
\begin{align}
\langle p',s'|T_a^{\mu\nu}(0)|p,s\rangle
&=
\overline u(p',s')
\Bigg[
A^a(t,\mu)\gamma^{(\mu}P^{\nu)}
\nonumber\\
&\quad+
B^a(t,\mu)
\frac{P^{(\mu}i\sigma^{\nu)\rho}\Delta_\rho}{2M}
\nonumber\\
&\quad+
D^a(t,\mu)
\frac{\Delta^\mu\Delta^\nu-g^{\mu\nu}\Delta^2}{4M}
\nonumber\\
&\quad+
\overline C^{\,a}(t,\mu)M g^{\mu\nu}
\Bigg]
u(p,s),
\label{eq:emt-decomposition}
\end{align}
where parentheses denote symmetrization with a factor $1/2$. The form factors associated with individual quark flavors and the gluon operator depend on the renormalization scale and scheme. This dependence cancels appropriately only in matrix elements of the complete conserved energy--momentum tensor.

The quark and gluon contributions are not separately conserved. Consequently, the term proportional to $\overline C^{\,a}(t,\mu)$ must not be removed by imposing conservation on each partonic contribution separately. Conservation of the total energy--momentum tensor instead implies
\begin{equation}
\sum_a\overline C^{\,a}(t,\mu)=0,
\label{eq:Cbar-total-zero}
\end{equation}
where the sum extends over all active quark flavors and the gluon contribution.

The local twist-two operator generated by the second Mellin moment is the symmetric-traceless part of the quark operator. Contracting Eq.~\eqref{eq:emt-decomposition} with two lightlike vectors removes the metric terms and gives
\begin{align}
\int_{-1}^{1}dx\,xH^q(x,\xi,t;\mu)
&=
A^q(t,\mu)+\xi^2D^q(t,\mu),
\label{eq:H-second-moment}\\
\int_{-1}^{1}dx\,xE^q(x,\xi,t;\mu)
&=
B^q(t,\mu)-\xi^2D^q(t,\mu).
\label{eq:E-second-moment}
\end{align}
For comparison with an alternative convention, we denote the coefficient
of the highest polynomial power by $\mathcal C^q(t)$. It is related to the
form factor used here by
\begin{equation}
D^q(t)=4\mathcal C^q(t).
\label{eq:D-C-convention}
\end{equation}
The calligraphic notation distinguishes $\mathcal C^q(t)$ from both the
normalization coefficient $C_q$ of the longitudinal distribution and the
nonconservation form factor $\overline C^{\,q}(t)$ appearing in
Eq.~\eqref{eq:emt-decomposition}.

At zero skewness, the adopted valence-only boundary condition gives
\begin{align}
A^q(t,\mu_0)
&=
\int_0^1dx\,xq_v(x)
\exp\!\left[t f_q(x)\right],
\label{eq:Aq-zero-skewness}\\
B^q(t,\mu_0)
&=
\int_0^1dx\,xe_v^q(x)
\exp\!\left[t g_q(x)\right].
\label{eq:Bq-zero-skewness}
\end{align}
Thus, the electromagnetic reconstruction determines the form factors $A^q$ and $B^q$ within the low-scale model. It does not determine $D^q$, because the $D$-term disappears at $\xi=0$, and it does not determine $\overline C^{\,q}$, which is absent from the symmetric-traceless twist-two projection.

The interpretation of these quantities should also be kept distinct. At $t=0$, $A^q(0,\mu)$ is the longitudinal momentum fraction carried by quarks and antiquarks of flavor $q$ at the scale $\mu$. It is not, by itself, a rest-frame mass fraction, a three-dimensional mass density, or a contribution to the QCD trace anomaly. Such interpretations require additional combinations of energy--momentum-tensor form factors and a specified spatial framework.


\subsection{Momentum and angular-momentum constraints}
\label{sec:emt-sum-rules}

Ji's relation identifies the kinetic total angular momentum carried
by quarks of flavor $q$ as~\cite{Ji1997}
\begin{equation}
J^q(\mu)
=
\frac{1}{2}
\left[
A^q(0,\mu)+B^q(0,\mu)
\right].
\label{eq:Ji-relation}
\end{equation}
The decomposition of $J^q$ into quark-spin and quark-orbital contributions requires, in addition, the helicity distribution or the corresponding axial matrix element. Electromagnetic information alone does not determine that separation.

For convenience, we also define the finite-$t$ combination
\begin{equation}
J^q(t,\mu)
=
\frac{1}{2}
\left[
A^q(t,\mu)+B^q(t,\mu)
\right].
\label{eq:Jq-finite-t}
\end{equation}
At $t\neq0$, this is a useful form-factor combination, but it should not automatically be interpreted as the Fourier transform of a unique spatial angular-momentum density. Such an interpretation depends on the reference frame, localization prescription, and choice of density.

For the complete conserved energy--momentum tensor, Poincar\'e symmetry gives
\begin{equation}
\sum_a A^a(0,\mu)=1,
\quad
\sum_a B^a(0,\mu)=0,
\quad
\sum_a J^a(\mu)=\frac{1}{2}.
\label{eq:total-emt-constraints}
\end{equation}
The first relation is the momentum sum rule, whereas the second expresses the vanishing total anomalous gravitomagnetic moment. The third follows from the first two.

In QCD, the sums in Eq.~\eqref{eq:total-emt-constraints} extend over all active quark flavors and the gluon contribution; antiquarks are already included in the quark matrix elements. In the present valence-only model, identifying the effective dressed $u$- and $d$-quark degrees of freedom with the complete proton
energy--momentum tensor constitutes a low-scale matching assumption. The corresponding target conditions are
\begin{equation}
A^{u+d}(0,\mu_0)=1,
\qquad
B^{u+d}(0,\mu_0)=0.
\label{eq:low-scale-emt-constraints}
\end{equation}
The first relation should be satisfied by the fixed longitudinal input of Eq.~\eqref{eq:valence-input}, whereas the second is imposed analytically through the Pauli-shape parameters. Neither condition follows from the electromagnetic normalizations $F_1^q(0)$ and $F_2^q(0)$.

\begin{table}[!t]
\caption{\label{tab:emt-zero-transfer}
Zero-transfer energy--momentum-tensor form factors and kinetic quark
angular momenta at the hadronic input scale \(\mu_0\). The last row
provides a numerical audit of the low-scale momentum and anomalous
gravitomagnetic-moment constraints.}
\begin{ruledtabular}
\begin{tabular}{cccc}
Contribution & $A^a(0)$ & $B^a(0)$ & $J^a$ \\
\hline
\tstrut
$u$   & 0.6880 &  0.5268 &  0.6074 \\
$d$   & 0.3125 & -0.5267 & -0.1071 \\
$u+d$ & 1.0005 &  0.0000 &  0.5003 \\
\end{tabular}
\end{ruledtabular}
\end{table}

Because the low-scale momentum condition is not imposed analytically through the longitudinal exponents, the fixed input gives
\begin{equation}
A^{u+d}(0,\mu_0)=1.0005,
\end{equation}
corresponding to a momentum-sum residual of approximately $5\times10^{-4}$. By contrast, $B^{u+d}(0,\mu_0)=0$ is imposed analytically through the derived parameter $\gamma_d$, leaving only a numerical integration residual below $10^{-4}$. The individual second moments are
\begin{align}
B^u(0) &= 0.526768,
&
B^d(0) &= -0.526722,
\nonumber\\
J^u &= 0.607384,
&
J^d &= -0.107111.
\label{eq:Jnumbers}
\end{align}
Consequently,
\begin{equation}
J^{u+d} = \frac{1}{2} \left[ A^{u+d}(0)+B^{u+d}(0) \right] = 0.5003,
\end{equation}
rather than exactly $1/2$ at the quoted numerical precision.

\begin{figure*}
\includegraphics[width=.90\textwidth]{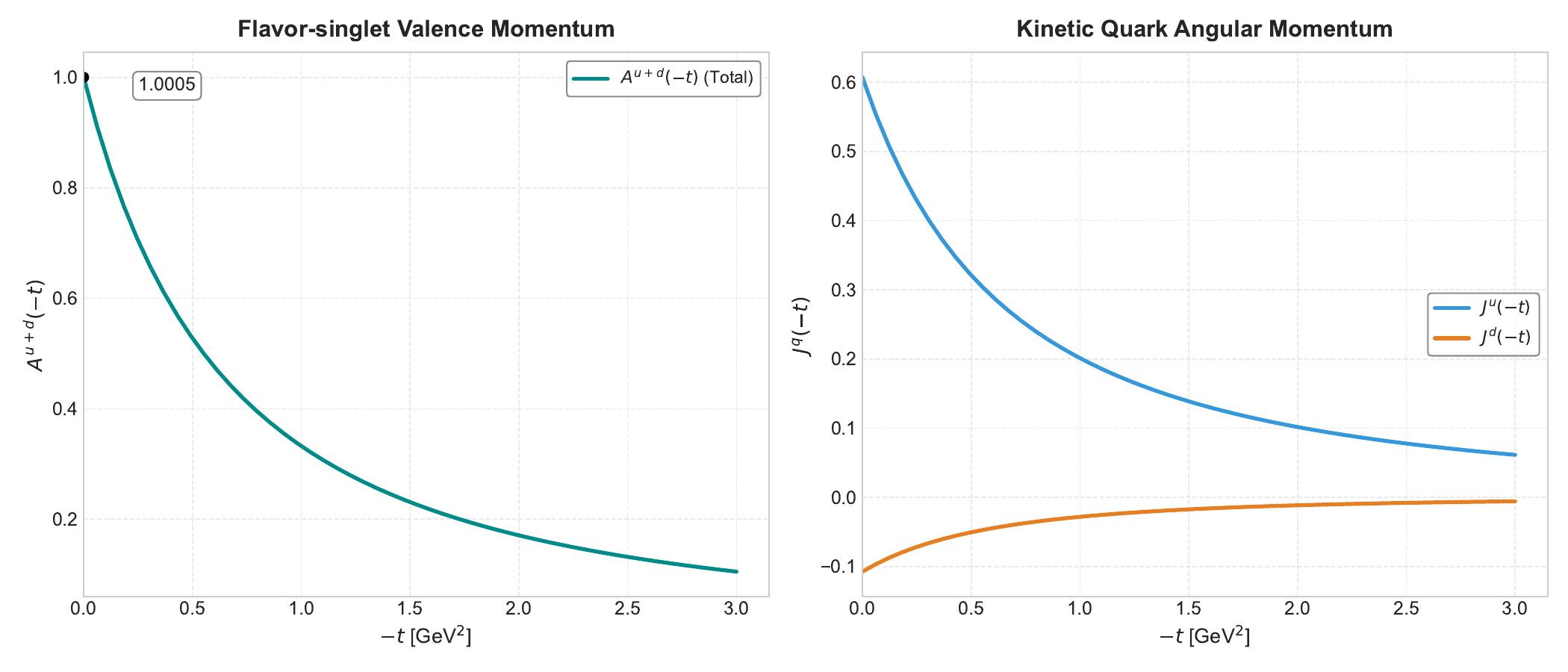}
\caption{\label{fig:emt-form-factors} Flavor-singlet valence momentum form factor $A^{u+d}(-t)$ (left) and the finite-$t$ combinations $J^q(-t)=[A^q(-t)+B^q(-t)]/2$ for $q=u,d$ (right), obtained from the constrained zero-skewness reconstruction. At $t=0$, the imposed condition $B^{u+d}(0)=0$, together with $A^{u+d}(0)=1.0005$, gives $J^{u+d}=0.5003$. For $t\neq0$, the functions $J^q(t)$ are useful form-factor combinations and are not interpreted here as unique spatial angular-momentum densities.}
\end{figure*}

The momentum transfer dependence of $A^q(t)$, $B^q(t)$, and $J^q(t)$ is a prediction of the fitted zero-skewness ansatz within its calibrated range. It is displayed in Fig.~\ref{fig:emt-form-factors}. The decrease of these form factors with increasing $-t$ reflects the $x$-dependent transverse profiles entering Eqs.~\eqref{eq:Aq-zero-skewness} and~\eqref{eq:Bq-zero-skewness}. The resulting curves remain scale-dependent flavor contributions and should not be compared directly with lattice-QCD results obtained at a different renormalization scale. Such a comparison requires perturbative evolution, compatible operator conventions, and the inclusion of sea-quark and gluon components.

\subsection{Information not fixed by electromagnetic data}
\label{sec:emt-undetermined}

The electromagnetic calibration constrains \(A^q(t)\) and \(B^q(t)\)
only through the adopted relation between the zero-skewness GPDs and
the fitted Dirac and Pauli form factors. It provides no direct
constraint on the \(D\)-form factor. The latter multiplies the
highest power of \(\xi\) in
Eqs.~\eqref{eq:H-second-moment} and
\eqref{eq:E-second-moment} and must be introduced independently when
the GPDs are extended to nonzero skewness.

Similarly, the form factor \(\overline C^{\,q}(t)\) cannot be obtained
from leading-twist GPD moments. For the complete conserved tensor its
sum over all partonic species vanishes, but the individual quark and
gluon contributions need not vanish. Consequently, the present
analysis does not provide a complete decomposition of the proton
energy, pressure, trace, or mass into separate partonic components.

These limitations are physically useful. They identify precisely which gravitational information is inherited from the electromagnetically constrained zero-skewness distributions and which information enters only through the independent nonzero-skewness and mechanical sectors developed below.


\section{Extension to nonzero skewness}
\label{sec:nonzero-skewness}

The electromagnetic form factors constrain only the zero-skewness GPDs introduced in Sec.~\ref{sec:zero-skewness}. Extending these distributions to $\xi\neq 0$ therefore requires additional model information. In this work we employ a factorized double-distribution ansatz, commonly referred to as the Radyushkin double-distribution ansatz (RDDA)~\cite{Radyushkin1997, PolyakovWeiss1999, Diehl2003, Belitsky2005}. This construction provides the correct GPD support and generates Mellin moments with the polynomial structure required by Lorentz covariance. It does not, however, determine the skewness dependence uniquely.

\subsection{Factorized double distributions}

For the positive-$\beta$ valence benchmark, we define
\begin{align}
F_H^q(\beta,\alpha,t)
&=
H_v^q(\beta,0,t)\,h_b(\beta,\alpha)
\nonumber\\
&=
q_v(\beta)\exp\!\left[t f_q(\beta)\right]
h_b(\beta,\alpha),
\label{eq:DD-H}
\\
F_E^q(\beta,\alpha,t)
&=
E_v^q(\beta,0,t)\,h_b(\beta,\alpha)
\nonumber\\
&=
e_v^q(\beta)\exp\!\left[t g_q(\beta)\right]
h_b(\beta,\alpha),
\label{eq:DD-E}
\end{align}
with
\begin{equation}
0\leq\beta\leq 1, \qquad -1+\beta\leq\alpha\leq 1-\beta .
\end{equation}
The profile function is taken to be
\begin{equation}
h_b(\beta,\alpha)
=
\frac{\Gamma(2b+2)}
     {2^{\,2b+1}\Gamma^2(b+1)}
\frac{\bigl[(1-\beta)^2-\alpha^2\bigr]^b}
     {(1-\beta)^{2b+1}},
\qquad b>-1 .
\label{eq:profile-hb}
\end{equation}
It obeys
\begin{equation}
\int_{-1+\beta}^{1-\beta}d\alpha\,
h_b(\beta,\alpha)=1,
\qquad
h_b(\beta,-\alpha)=h_b(\beta,\alpha).
\label{eq:profile-properties}
\end{equation}
Consequently, the construction reproduces the fitted zero-skewness GPDs:
\begin{align}
&
H_{\rm DD}^q(x,0,t)=H_v^q(x,0,t), \qquad 0<x<1, \\
&
E_{\rm DD}^q(x,0,t)=E_v^q(x,0,t), \qquad 0<x<1 .
\label{eq:DD-zero-skewness}
\end{align}

The parameter $b$ controls the width of the double distribution in the $\alpha$ direction and hence the skewness dependence. Increasing $b$ concentrates the profile around $\alpha=0$, with $h_b(\beta,\alpha)\to\delta(\alpha)$ in the formal $b\to\infty$ limit. The value of $b$ is not determined by the number of valence quarks or by the electromagnetic form-factor fit. It is an independent model parameter. For simplicity, we use the same profile parameter for $H$ and $E$ and for both valence flavors. Allowing separate profile parameters would introduce additional model freedom that cannot be constrained by the elastic electromagnetic information considered here. We use $b=3$ as the central benchmark and compare it with $b=1$ and $b=8$ to display the corresponding model sensitivity. For the central choice,
\begin{equation}
h_3(\beta,\alpha) = \frac{35}{32} \frac{\bigl[(1-\beta)^2-\alpha^2\bigr]^3}{(1-\beta)^7}.
\label{eq:profile-b3}
\end{equation}

The double-distribution contributions to the GPDs are
\begin{align}
H_{\rm DD}^q(x,\xi,t) &= \int_0^1d\beta \int_{-1+\beta}^{1-\beta}d\alpha\, \delta(x-\beta-\xi\alpha) F_H^q(\beta,\alpha,t),
\label{eq:H-DD}
\\
E_{\rm DD}^q(x,\xi,t) &= \int_0^1d\beta \int_{-1+\beta}^{1-\beta}d\alpha\,
\delta(x-\beta-\xi\alpha) F_E^q(\beta,\alpha,t).
\label{eq:E-DD}
\end{align}

Equations~\eqref{eq:H-DD} and~\eqref{eq:E-DD} use only positive values of $\beta$. They therefore implement the absence of an independent forward antiquark input at the hadronic matching scale adopted in this work. A complete quark--antiquark GPD representation would require both signs of $\beta$, together with the appropriate antiquark conventions. The positive-$\beta$ construction may nevertheless generate support at negative $x$ inside the ERBL region, $|x|<\xi$, when $\xi\neq0$.

The double-distribution representation guarantees the correct support and polynomial Mellin moments. These properties should not be confused with GPD positivity: the factorized ansatz does not, by itself, establish the general positivity inequalities at nonzero skewness. Accordingly, the present construction should be regarded as a polynomiality-preserving extension of the constrained zero-skewness input, rather than as a proof of positivity for arbitrary $\xi$.


\subsection{One-dimensional representation}

For numerical calculations it is convenient to perform the $\alpha$ integration analytically. We take $0<\xi<1$. For the unpolarized GPDs considered here, time-reversal invariance implies $H^q(x,-\xi,t)=H^q(x,\xi,t)$ and $E^q(x,-\xi,t)=E^q(x,\xi,t)$, so that negative skewness need not be treated separately. The delta function fixes
\begin{equation}
 \alpha=\frac{x-\beta}{\xi}
\end{equation}
and produces a Jacobian $1/\xi$. The spectral support condition becomes
\begin{equation}
|x-\beta|\leq\xi(1-\beta).
\label{eq:DD-support-condition}
\end{equation}
Solving this inequality gives
\begin{align}
\beta_{\min}
&=
\max\left(0,\frac{x-\xi}{1-\xi}\right),
\label{eq:beta-min}
\\
\beta_{\max}
&=
\min\left(1,\frac{x+\xi}{1+\xi}\right).
\label{eq:beta-max}
\end{align}
If $\beta_{\min}<\beta_{\max}$, the GPDs can be evaluated as
\begin{align}
H_{\rm DD}^q(x,\xi,t)
&=
\frac{1}{\xi}
\int_{\beta_{\min}}^{\beta_{\max}}d\beta\,
F_H^q\left(
\beta,\frac{x-\beta}{\xi},t
\right),
\label{eq:H-DD-one-dimensional}
\\
E_{\rm DD}^q(x,\xi,t)
&=
\frac{1}{\xi}
\int_{\beta_{\min}}^{\beta_{\max}}d\beta\,
F_E^q\left(
\beta,\frac{x-\beta}{\xi},t
\right).
\label{eq:E-DD-one-dimensional}
\end{align}
The result is zero when the integration interval is empty. This form is numerically preferable to approximating the delta function by a sharply peaked finite-width function.

With the positive-$\beta$ input, $H_{\rm DD}^q=E_{\rm DD}^q=0$ for $x<-\xi$. In the ERBL region, $-\xi<x<\xi$, the lower integration limit is zero, and the GPD may be nonzero even at negative $x$. In the DGLAP region, $x>\xi$, both endpoints depend on $x$. At $\xi=0$, the zero-skewness expressions should be used directly rather than evaluating Eqs.~\eqref{eq:H-DD-one-dimensional} and~\eqref{eq:E-DD-one-dimensional} with an artificially small value of $\xi$.

The ERBL contribution generated in this way is the nonzero-skewness completion implied by the chosen double-distribution ansatz. It should not be interpreted as a uniquely determined sea-quark distribution or as evidence for a particular higher-Fock-state overlap. The double-distribution construction and a truncation of the proton light-front wave function constitute different model assumptions.


\begin{figure*}[!t]
\includegraphics[width=.90\textwidth]{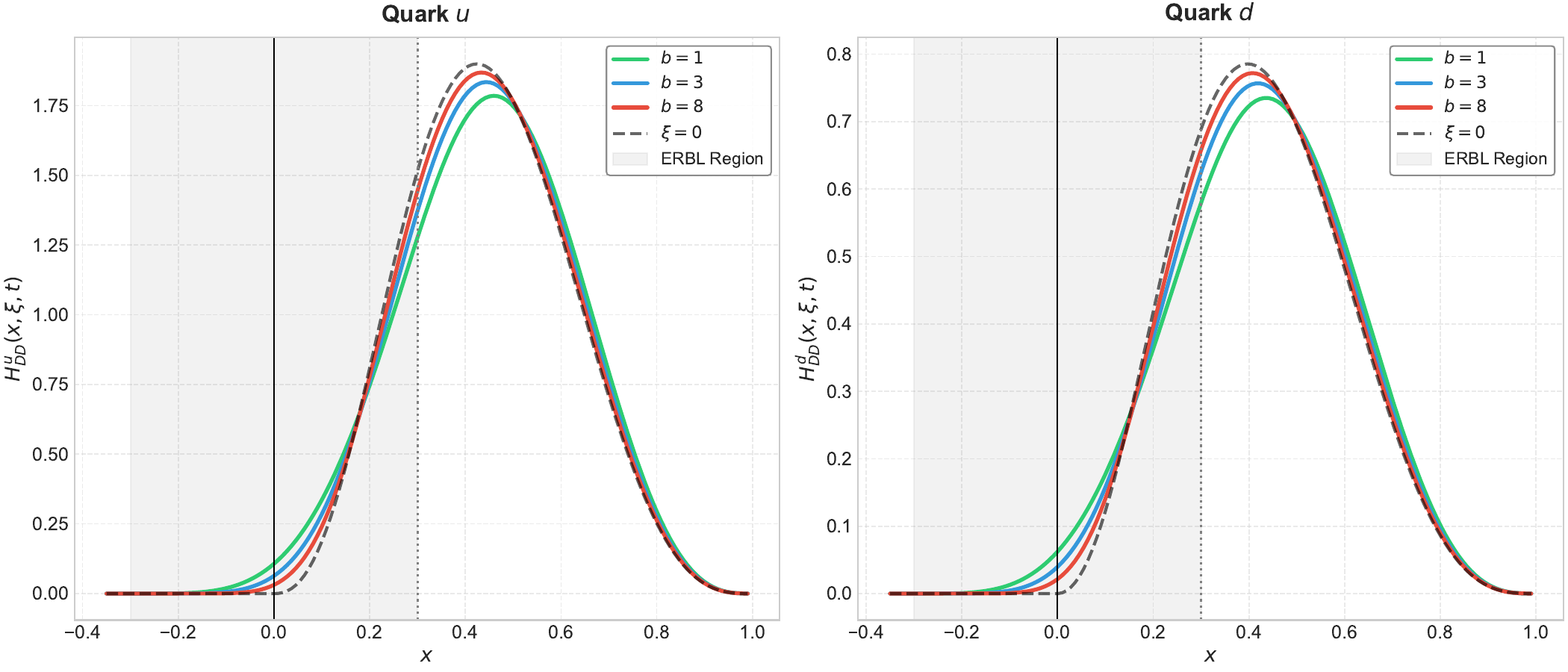}
\caption{\label{fig:DD-profile-comparison} Sensitivity of the DD-only quark GPDs to the profile parameter $b$ at $\xi=0.3$ and $t=-0.5\,\mathrm{GeV}^2$, evaluated with the positivity-constrained zero-skewness parameters. The dashed curve shows the corresponding zero-skewness input evaluated at the same value of $t$, and the shaded area indicates the ERBL region. All displayed values of $b$ give the same electromagnetic moment $\int dx\,H_{\rm DD}^q$ and spin-two moment $\int dx\,xH_{\rm DD}^q$; their differences therefore illustrate the model freedom in the ERBL region and in higher Mellin moments.}
\end{figure*}

\subsection{Polynomiality and the role of the profile}

Using Eq.~\eqref{eq:H-DD}, the $n$th Mellin moment becomes
\begin{align}
\int_{-1}^{1}dx\,x^n H_{\rm DD}^q(x,\xi,t)
&=
\sum_{k=0}^{n} {n\choose k}\xi^k \nonumber \\
&
\hspace{-2.00cm} \times \int_0^1 d\beta \int_{-1+\beta}^{1-\beta}d\alpha\, \beta^{\,n-k}\alpha^k F_H^q(\beta,\alpha,t).
\label{eq:DD-polynomiality}
\end{align}
Because the profile is even in $\alpha$, all odd-$k$ terms vanish. The
result is therefore an even polynomial in $\xi$ of degree not larger
than $n$. For odd $n$, the highest power allowed by Lorentz covariance,
$\xi^{n+1}$, is absent from the present double-distribution
contribution. Its coefficient is supplied by the independent
$D$-term introduced in Sec.~\ref{sec:D-term}.

For $n=0$, normalization of the profile gives
\begin{align}
\int_{-1}^{1}dx\, H_{\rm DD}^q(x,\xi,t) = F_1^q(t), \nonumber \\
\int_{-1}^{1}dx\, E_{\rm DD}^q(x,\xi,t) = F_2^q(t).
\label{eq:DD-electromagnetic-moments}    
\end{align}
Thus the extension preserves the electromagnetic form factors and does not introduce a spurious skewness dependence into them.

For $n=1$, the term linear in $\alpha$ vanishes and one obtains
\begin{align}
\int_{-1}^{1}dx\,xH_{\rm DD}^q(x,\xi,t)
&=A^q(t),
\label{eq:DD-second-moment-H}
\\
\int_{-1}^{1}dx\,xE_{\rm DD}^q(x,\xi,t)
&=B^q(t).
\label{eq:DD-second-moment-E}
\end{align}
The DD-only construction therefore corresponds to a vanishing coefficient of the $\xi^2$ term in these moments. This is a permissible special case of polynomiality, not a violation of Lorentz covariance. The independent $D$-term restores the most general allowed $\xi^2$ contribution.

The first profile-dependent contribution appears for $n=2$. For $H^q$,
\begin{align}
\int_{-1}^{1}dx\,x^2H_{\rm DD}^q(x,\xi,t)
&= \int_0^1d\beta\, q_v(\beta)e^{t f_q(\beta)} \nonumber \\
&
\times \left[ \beta^2+ \xi^2\frac{(1-\beta)^2}{2b+3} \right].
\label{eq:DD-third-moment}
\end{align}
An analogous expression holds for $E^q$ after replacing $q_v e^{t f_q}$ by $e_v^q e^{t g_q}$. This result makes explicit that the parameter $b$ controls information not contained in either the electromagnetic moments or the spin-two form factors.

Figure~\ref{fig:DD-profile-comparison} compares the choices $b=1$, $3$, and $8$ at fixed $\xi$ and $t$. All three curves reproduce the same fitted zero-skewness GPD, electromagnetic form factor, and spin-two form factor. Their differences in the ERBL region and near $x=\xi$ quantify the sensitivity to the assumed profile; they are not statistical uncertainties derived from the electromagnetic fit. Moreover, the nonzero-skewness extension inherits all limitations of its zero-skewness input. Polynomiality alone cannot repair a violation of a separate theoretical constraint.


\section{Independent completion by the $D$-term}
\label{sec:D-term}

The factorized double-distribution construction of Sec.~\ref{sec:nonzero-skewness} generates polynomial Mellin moments but does not, in general, supply the highest power of $\xi$ allowed by Lorentz covariance. For the unpolarized quark GPDs, this missing contribution can be introduced through an independent $D$-term~\cite{PolyakovWeiss1999, Diehl2003, Belitsky2005}. Its normalization and factorization-scale dependence must be specified independently of the electromagnetic fit.

\subsection{Definition and normalization}

To distinguish the function supported in the ERBL region from the corresponding energy--momentum-tensor form factor, we denote the former by $\mathcal D^q(z,t;\mu)$. We define
\begin{align}
H^q(x,\xi,t;\mu) &= H_{\rm DD}^q(x,\xi,t;\mu) + H_D^q(x,\xi,t;\mu),
\label{eq:H-complete}
\\
E^q(x,\xi,t;\mu) &= E_{\rm DD}^q(x,\xi,t;\mu) - H_D^q(x,\xi,t;\mu),
\label{eq:E-complete}
\end{align}
where
\begin{equation}
H_D^q(x,\xi,t;\mu) = \operatorname{sgn}(\xi)\, \theta\bigl(|\xi|-|x|\bigr) \mathcal D^q\left(\frac{x}{\xi},t;\mu\right).
\label{eq:D-term-contribution}
\end{equation}
For the unpolarized vector GPD, the function $\mathcal D^q(z,t;\mu)$ is odd in $z$. The factor $\operatorname{sgn}(\xi)$ then ensures that the GPD is even under $\xi\to-\xi$, as required by time-reversal invariance.

The same $D$-term enters $H^q$ and $E^q$ with opposite signs. Consequently,
\begin{align}
H^q(x,\xi,t;\mu)+E^q(x,\xi,t;\mu) &= H_{\rm DD}^q(x,\xi,t;\mu) \nonumber \\
&
+ E_{\rm DD}^q(x,\xi,t;\mu),
\label{eq:D-cancellation}
\end{align}
so that the $D$-term cancels from the combination entering Ji's relation. Adding it to $H^q$ but not to $E^q$ would generate an unphysical $\xi$ dependence in the second moment of $H^q+E^q$.

We use the Gegenbauer expansion
\begin{equation}
\mathcal D^q(z,t;\mu) = (1-z^2) \sum_{\substack{n=1\\n\ {\rm odd}}}^{\infty}
d_n^q(t;\mu)\, C_n^{3/2}(z).
\label{eq:D-Gegenbauer}
\end{equation}
The first two polynomials are
\begin{equation}
C_1^{3/2}(z)=3z,
\qquad
C_3^{3/2}(z)=\frac52(7z^3-3z).
\end{equation}
In the conventions adopted in Sec.~\ref{sec:gravitational}, the quark energy--momentum-tensor form factor is
\begin{align}
D^q(t;\mu)
&=
\int_{-1}^{1}dz\,z\, \mathcal D^q(z,t;\mu) \nonumber\\
&=
\frac45d_1^q(t;\mu).
\label{eq:D-form-factor-d1}
\end{align}
Higher Gegenbauer coefficients do not contribute to this spin-two form factor because of orthogonality. In particular, if only $d_1^q$ is retained,
\begin{equation}
\mathcal D^q(z,t;\mu) = \frac{15}{4}D^q(t;\mu)\, z(1-z^2).
\label{eq:D-leading-harmonic}
\end{equation}

For odd $m$, direct substitution of Eq.~\eqref{eq:D-term-contribution} gives
\begin{equation}
\int_{-1}^{1}dx\,x^m H_D^q(x,\xi,t;\mu) = \xi^{m+1} \int_{-1}^{1}dz\,z^m \mathcal D^q(z,t;\mu),
\label{eq:D-Mellin}
\end{equation}
whereas the integral vanishes for even $m$. For $m=1$, one recovers
\begin{align}
\int_{-1}^{1}dx\,xH^q(x,\xi,t;\mu)
&=
A^q(t;\mu)+\xi^2D^q(t;\mu),
\label{eq:H-second-moment-complete}
\\
\int_{-1}^{1}dx\,xE^q(x,\xi,t;\mu)
&=
B^q(t;\mu)-\xi^2D^q(t;\mu).
\label{eq:E-second-moment-complete}
\end{align}
The electromagnetic moments remain unchanged because the $D$-term is odd in $z$:
\begin{equation}
\int_{-1}^{1}dx\,H_D^q(x,\xi,t;\mu)=0.
\end{equation}


\begin{figure*}[!t]
\includegraphics[width=0.95\textwidth]{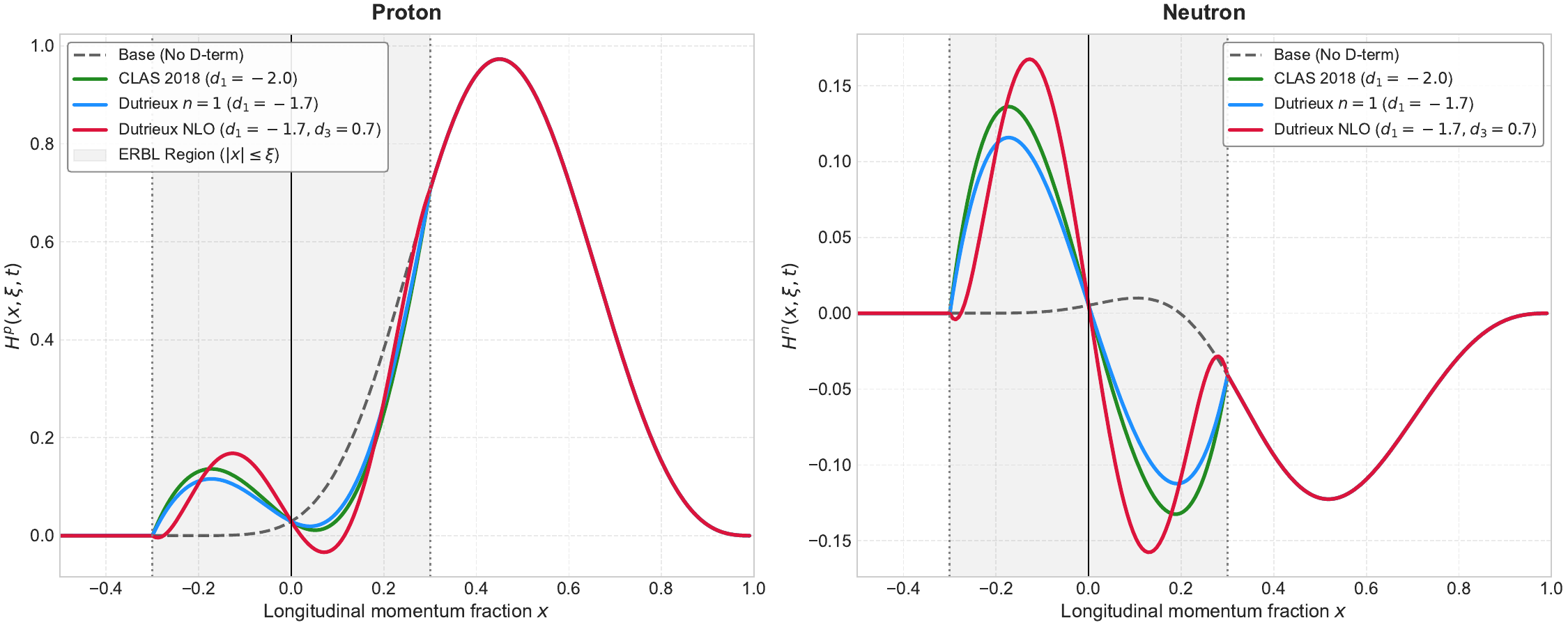}
\caption{\label{fig:D-term-benchmarks} Illustration of the effect of phenomenologically motivated $D$-term scenarios on the DD baseline at the kinematics specified in the text. The dashed curve shows the DD-only result obtained from the positivity-constrained zero-skewness input. The solid curves include the ERBL-supported $D$-term for the benchmark coefficient pairs $(d_1,d_3)=(-2.0,0)$, $(-1.7,0)$, and $(-1.7,0.7)$. The latter two pairs use central values reported in the NLO analysis of Ref.~\cite{Dutrieux2025}, but their use here does not constitute an NLO calculation. The shaded area marks the ERBL region $|x|\leq\xi$.}
\end{figure*}

\subsection{Illustrative benchmark scenarios}

The electromagnetic calibration performed in this work provides no constraint on
$\mathcal D^q(z,t;\mu)$ or $D^q(t;\mu)$. Any numerical $D$-term introduced below must therefore be regarded as an independent input.

For illustration, we consider truncated expansions containing the coefficient pairs
\begin{equation}
(d_1,d_3)=(-2.0,0),\quad (-1.7,0), \quad (-1.7,0.7).
\label{eq:D-benchmarks}
\end{equation}
These values are motivated by central values appearing in phenomenological analyses of DVCS data~\cite{Burkert2018, Dutrieux2025}, but we do not interpret them as precise determinations.

For a transparent analytic benchmark, we assign a common tripole dependence,
\begin{equation}
d_n(t)= \frac{d_n(0)}{\left(1-t/m_D^2\right)^3}, \qquad m_D=0.8~{\rm GeV}.
\label{eq:D-tripole}
\end{equation}
This transfer dependence is an additional model choice and is not determined by the elastic electromagnetic fit. Likewise, applying a light-flavor-singlet coefficient to separate proton and neutron GPDs requires a flavor decomposition. Any such decomposition must be stated explicitly.

Figure~\ref{fig:D-term-benchmarks} illustrates the effect of the three benchmark scenarios in Eq.~\eqref{eq:D-benchmarks} on the complete GPD at fixed $\xi$ and $t$: $H^p(x,\xi,t) = (2/3)H^u(x,\xi,t) - (1/3)H^d(x,\xi,t)$ and $H^n(x,\xi,t) = (2/3)H^d(x,\xi,t) - (1/3)H^u(x,\xi,t)$. Since the $D$-term has support only in the ERBL region, all scenarios coincide with the DD baseline for $|x|>\xi$, whereas they generate distinct modifications for $|x|\leq\xi$. The comparison between the $(d_1,d_3)=(-1.7,0)$ and $(-1.7,0.7)$ scenarios shows directly how a higher Gegenbauer harmonic can modify the shape of the GPD in the ERBL region without changing the spin-two form factor $D(t)$. These curves are illustrative benchmarks and should not be interpreted as flavor-separated extractions from the electromagnetic calibration.


\subsection{Higher harmonics and shadow directions}

Equation~(\ref{eq:D-form-factor-d1}) shows that $d_3^q$, and all higher Gegenbauer coefficients, are invisible to the spin-two form factor $D^q(t)$. Changing $d_3^q$ at fixed $d_1^q$ modifies the GPD inside the ERBL region and changes higher Mellin moments, but it leaves the pressure and shear distributions constructed solely from the same spin-two form factor $D^q(t)$ unchanged.

This moment-space insensitivity is distinct from the information contained in a DVCS convolution. At leading order, the quark contribution to the vector Compton form factor may be written schematically as
\begin{align}
\mathcal H(\xi,t) &= \sum_q e_q^2 \int_{-1}^{1}dx\, H^q(x,\xi,t) \nonumber \\
&
\times \left[ \frac{1}{\xi-x-i0} - \frac{1}{\xi+x-i0} \right].
\label{eq:LO-CFF}
\end{align}
The $D$-term contribution is real and independent of $\xi$:
\begin{equation}
\mathcal H_D(t) = 4\sum_q e_q^2 \sum_{\substack{n=1\\n\ {\rm odd}}}^{\infty} d_n^q(t)
\label{eq:subtraction-constant}
\end{equation}
in the conventions of Eq.~\eqref{eq:D-Gegenbauer}. It therefore constrains a different linear combination of Gegenbauer coefficients from the spin-two form factor. For example, a variation satisfying
\begin{equation}
\delta d_1^q(t)=-\delta d_3^q(t)
\end{equation}
leaves the leading-order subtraction constant unchanged at a fixed scale while modifying \(D^q(t)\). Such directions are commonly called shadow $D$-terms~\cite{Dutrieux2025}.

Dispersion relations relate the real and imaginary parts of the Compton amplitude and provide access to its subtraction constant~\cite{PasquiniPolyakovVanderhaeghen2014}. They do not, without additional assumptions or scale-evolution information, uniquely reconstruct every Gegenbauer coefficient of $\mathcal D^q(z,t)$. Moreover, a genuine NLO analysis requires NLO coefficient functions, quark--gluon evolution, a specified factorization scheme, and consistent quark and gluon inputs. Retaining both $d_1$ and $d_3$ is not by itself an NLO treatment.


\subsection{Numerical polynomiality test}

As a numerical validation, we evaluate the complete GPDs over $-1\leq x\leq1$, splitting the integration at $x=-\xi$, $x=0$, and $x=\xi$. The resulting moments
are compared with the independently calculated zero-skewness form factors:
\begin{align}
\int_{-1}^{1}dx\,xH^q(x,\xi,t)
&\stackrel{\rm num.}{=}
A^q(t)+\xi^2D^q(t),
\\
\int_{-1}^{1}dx\,xE^q(x,\xi,t)
&\stackrel{\rm num.}{=}
B^q(t)-\xi^2D^q(t).
\end{align}
At $t=-0.5~{\rm GeV}^2$, $b=3$, and $0.05\leq\xi\leq0.35$, the maximum absolute numerical discrepancy is below $2\times10^{-10}$.

\begin{figure}[!t]
\includegraphics[width=0.48\textwidth]{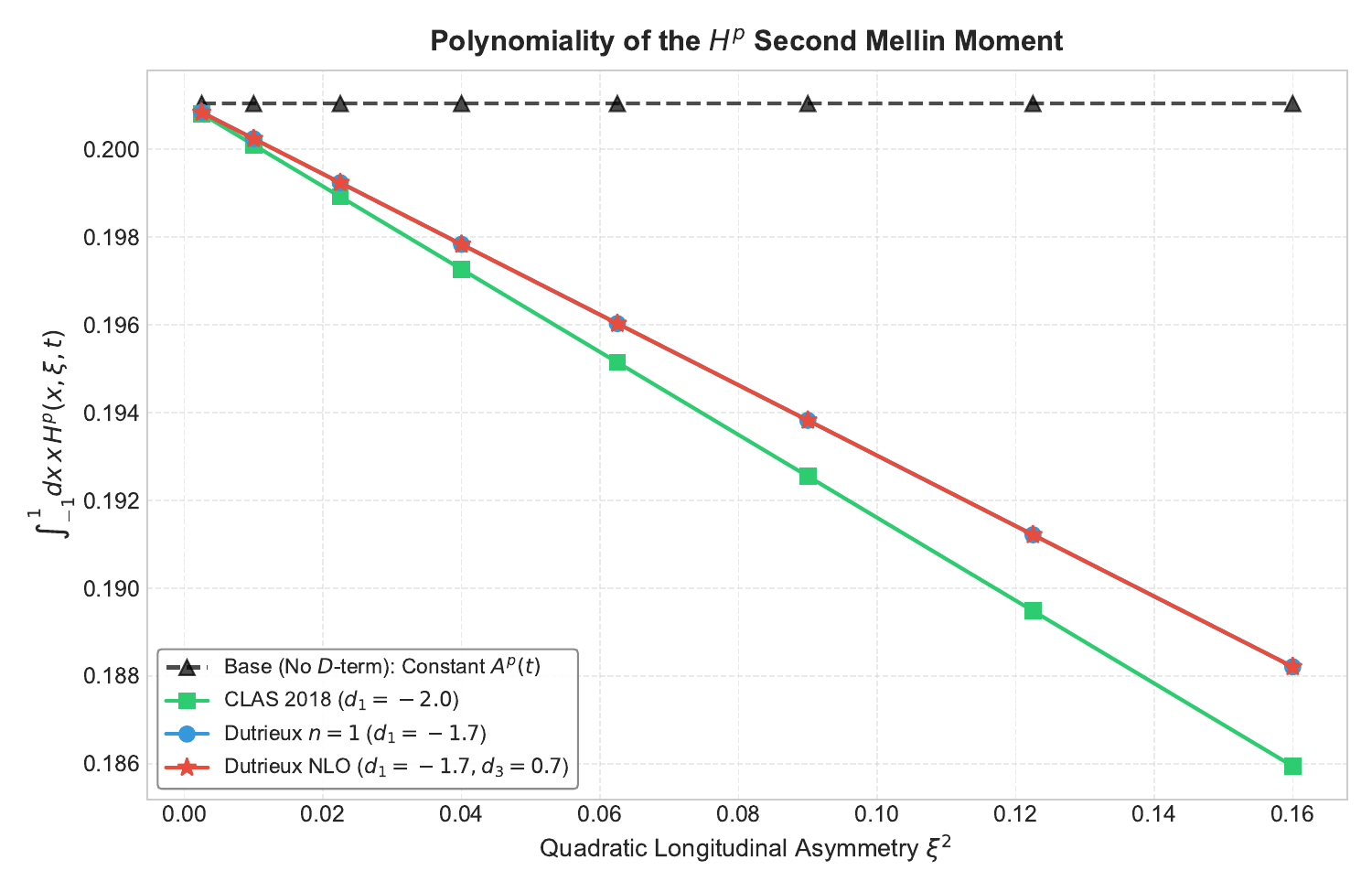}
\caption{\label{fig:polynomiality-test} Numerical verification of polynomiality through the spin-two moment $\int_{-1}^{1}dx\,xH(x,\xi,t)$ at $t=-0.5\,\mathrm{GeV}^2$. The DD-only baseline is independent of $\xi$. Adding the $n=1$ Gegenbauer harmonic generates the required $\xi^2D(t)$ contribution. Adding the $n=3$ harmonic at fixed $d_1$ leaves this moment unchanged, providing a numerical check of the orthogonality of the higher Gegenbauer harmonics to the spin-two form factor. The curves are illustrative benchmark scenarios and should not be interpreted as a fit to DVCS data.}
\end{figure}

The result of this test is displayed in Fig.~\ref{fig:polynomiality-test} for the complete GPD, \emph{i.e.} $H^p(x,\xi,t) = (2/3)H^u(x,\xi,t) - (1/3)H^d(x,\xi,t)$ and $H^n(x,\xi,t) = (2/3)H^d(x,\xi,t) - (1/3)H^u(x,\xi,t)$. The DD-only result is independent of $\xi$, as expected from Eq.~\eqref{eq:DD-second-moment-H}. Including the leading $n=1$ Gegenbauer harmonic produces the quadratic dependence $A(t)+\xi^2D(t)$. The curves obtained with $(d_1,d_3)=(-1.7,0)$ and $(-1.7,0.7)$ coincide because the $n=3$ harmonic is orthogonal to the spin-two moment. The figure therefore provides a simultaneous numerical check of the $D$-term normalization, the ERBL support, and the polynomiality relation.

The opposite slopes of the $H^q$ and $E^q$ moments provide a sensitive simultaneous check of the integration limits, the Jacobian, the sign convention, and the normalization of the $D$-term. The absence of a contribution from $d_3^q$ to these moments provides an additional check of Gegenbauer orthogonality. Polynomiality is, however, a structural consistency condition rather than evidence that a particular $D$-term parametrization reproduces DVCS data.

\section{Illustrative mechanical distributions}
\label{sec:mechanical-distributions}

The electromagnetic calibration performed in this work does not determine the mechanical properties of the proton. In particular, the $D$-term is absent from the electromagnetic form-factor sum rules and must be supplied independently. Furthermore, even if a quark contribution $D^q(t)$ is specified, it cannot in general be identified with the form factor of the complete conserved energy--momentum tensor. Such an identification requires a consistent treatment of all active quark flavors and gluons.

The purpose of this section is therefore limited. We use an illustrative model for the total conserved form factor $D(t)$ to demonstrate explicitly how a momentum-space $D$-term generates static pressure and shear distributions in the conventional Breit-frame representation.


\subsection{Static Breit-frame convention}

For a spherically symmetric system, define the three-dimensional Fourier transform
\begin{equation}
\widetilde D(r) = \int\frac{d^3\boldsymbol{\Delta}}{(2\pi)^3}\, e^{-i\boldsymbol{\Delta}\cdot\boldsymbol{r}}\, D(-\boldsymbol{\Delta}^2).
\label{eq:D-Fourier}
\end{equation}
The static spatial stress tensor is decomposed as
\begin{equation}
T^{ij}(\boldsymbol r)
=
\left( \frac{r^ir^j}{r^2}-\frac{\delta^{ij}}{3} \right) s(r) + \delta^{ij} p(r),
\label{eq:stress-decomposition}
\end{equation}
where $p(r)$ is the isotropic pressure and $s(r)$ is the shear-force distribution. In the convention adopted here~\cite{PolyakovSchweitzer2018},
\begin{align}
p(r) &= \frac{1}{6M} \frac{1}{r^2}\frac{d}{dr} \left[ r^2\frac{d\widetilde D(r)}{dr} \right],
\label{eq:pressure-from-D}
\\
s(r) &= -\frac{1}{4M} r\frac{d}{dr} \left[ \frac{1}{r}\frac{d\widetilde D(r)}{dr} \right].
\label{eq:shear-from-D}
\end{align}

These three-dimensional functions are defined in a static Breit-frame construction. They should not be interpreted as exact frame-independent probability densities of a relativistic proton. Relativistic localization and two-dimensional light-front stress distributions are distinct constructions~\cite{LorceMoutardeTrawinski2019}.

Conservation of the complete energy--momentum tensor
implies
\begin{equation}
\frac{2}{3}\frac{ds(r)}{dr} + \frac{2s(r)}{r} + \frac{dp(r)}{dr} = 0.
\label{eq:stress-conservation}
\end{equation}
The radial, or normal, pressure is
\begin{equation}
p_r(r)=p(r)+\frac{2}{3}s(r).
\label{eq:radial-pressure}
\end{equation}
Provided the corresponding surface term vanishes, EMT
conservation leads to the von Laue condition
\begin{equation}
\int_0^\infty dr\,r^2p(r)=0.
\label{eq:von-Laue}
\end{equation}
This relation expresses the cancellation between the repulsive and attractive regions of the isotropic pressure. It is a necessary equilibrium condition in the static construction, although its satisfaction alone does not establish every possible criterion for mechanical stability.

For a separately defined quark or gluon tensor, the nonconservation form factor $\overline C^a(t)$ can contribute to the isotropic stress. An individual partonic subsystem therefore need not satisfy the same isolated-equilibrium condition as the complete proton. For this reason, a charge-weighted or quark-only $D$-term should not be interpreted as the form factor of the complete conserved stress tensor.


\begin{figure*}[!t]
\includegraphics[width=.90\textwidth]{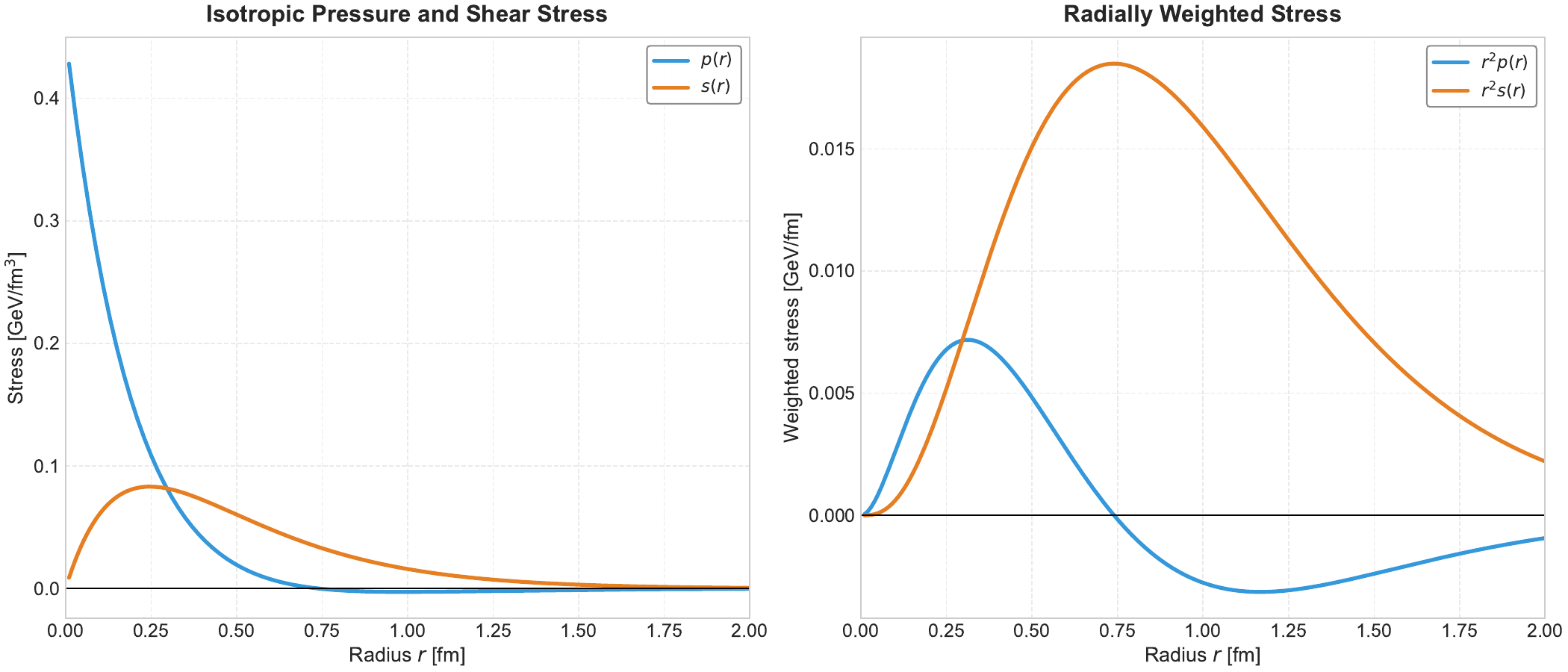}
\caption{\label{fig:tripole-stresses} Spatial stress distributions generated by the illustrative static tripole model. The left panel shows the isotropic pressure $p(r)$ and the shear-force distribution $s(r)$, while the right panel shows $r^2p(r)$ and $r^2s(r)$, for $D_0=-2$ and $m_D=0.8\,\mathrm{GeV}$. The pressure changes sign at $r_0=3/m_D$, and the positive inner contribution to $\int dr\,r^2p(r)$ is exactly balanced by the negative outer contribution, thereby satisfying the von Laue condition. The results refer to the independently specified total tripole form factor and not to any of the flavor-dependent $D$-term benchmarks considered in Sec.~\ref{sec:D-term}.}
\end{figure*}

\subsection{Analytic tripole benchmark}

Rather than attempting an extraction of the total proton stress from the electromagnetic fit, we consider the illustrative tripole form
\begin{equation}
D(t) = \frac{D_0}{\left(1-t/m_D^2\right)^3}.
\label{eq:D-tripole-mechanical}
\end{equation}
Here $D_0$ denotes the value of the total illustrative form factor at $t=0$. It is not the Gegenbauer coefficient $d_1^q$ of an individual quark flavor. Likewise, $m_D$ is an independent mechanical scale and is not determined by the electromagnetic calibration.

The tripole is used because it provides an analytically tractable and ultraviolet-convergent benchmark with finite stress distributions at the origin. Its three-dimensional Fourier transform is
\begin{equation}
\widetilde D(r) = \frac{D_0m_D^3}{32\pi}(1+m_Dr)e^{-m_Dr}.
\label{eq:D-tripole-transform}
\end{equation}
Writing $y=m_Dr$, Eqs.~\eqref{eq:pressure-from-D} and~\eqref{eq:shear-from-D} give
\begin{align}
p(r) &= \frac{D_0m_D^5}{192\pi M} (y-3)e^{-y},
\label{eq:tripole-pressure}
\\
s(r) &= -\frac{D_0m_D^5}{128\pi M} ye^{-y},
\label{eq:tripole-shear}
\\
p_r(r) &= -\frac{D_0m_D^5}{64\pi M} e^{-y}.
\label{eq:tripole-radial-pressure}
\end{align}

For $D_0<0$, the central pressure is positive, $s(r)\geq0$, and the radial pressure is positive for all finite $r$. The isotropic pressure changes sign at
\begin{equation}
r_0=\frac{3}{m_D}.
\label{eq:pressure-zero}
\end{equation}
The von Laue condition follows explicitly from
\begin{equation}
\int_0^\infty dy\,y^2(y-3)e^{-y}=0.
\label{eq:tripole-von-Laue}
\end{equation}
Thus, the positive inner pressure is exactly balanced by the negative outer pressure.

The distributions are regular at the origin:
\begin{equation}
p(0)= -\frac{D_0m_D^5}{64\pi M}, \qquad s(0)=0, \qquad \lim_{r\to0}r^2p(r)=0.
\label{eq:tripole-origin}
\end{equation}
For the illustrative values \(D_0=-2\) and
\(m_D=0.8~\mathrm{GeV}\), one obtains
\begin{equation}
p(0)=0.4521~\mathrm{GeV/fm^3},
\qquad
r_0=0.7400~\mathrm{fm}.
\label{eq:tripole-numerical-values}
\end{equation}
The conversion from natural units, $\mathrm{GeV}^4$, to $\mathrm{GeV/fm^3}$ is performed by dividing by $(\hbar c)^3$. These values characterize the chosen analytical benchmark and are not a new phenomenological extraction of the proton pressure.

A mechanical mean-square radius may be defined by weighting the
position with the positive radial pressure,
\begin{equation}
\langle r^2\rangle_{\rm mech} = \frac{\displaystyle \int d^3\boldsymbol r\,r^2p_r(r)} {\displaystyle \int d^3\boldsymbol r\,p_r(r)} = \frac{12}{m_D^2}.
\label{eq:mechanical-radius}
\end{equation}
For $m_D=0.8~\mathrm{GeV}$, this gives
\begin{equation}
\sqrt{\langle r^2\rangle_{\rm mech}} = 0.8545~\mathrm{fm}.
\end{equation}
This pressure-weighted definition should be distinguished from the slope parameter associated with the form factor,
\begin{equation}
\langle r^2\rangle_{D,\mathrm{slope}} \equiv \frac{6D'(0)}{D(0)} = \frac{18}{m_D^2}.
\label{eq:D-slope-radius}
\end{equation}
The two quantities characterize different aspects of the same tripole form factor and need not coincide. Any quoted mechanical radius should therefore be accompanied by its precise definition.

Figure~\ref{fig:tripole-stresses} displays the isotropic pressure $p(r)$ and shear-force distribution $s(r)$, together with their radially weighted forms. The sign change of $r^2p(r)$ separates the repulsive inner region from the attractive outer region, and the corresponding positive and negative contributions cancel exactly in accordance with Eq.~\eqref{eq:tripole-von-Laue}. For $D_0<0$, $r^2s(r)$ is nonnegative, while the radial pressure $p_r(r)=p(r)+2s(r)/3$ remains positive for every finite $r$.


\section{Conclusions and outlook}
\label{sec:conclusions}

We have developed a constrained phenomenological reconstruction of the unpolarized, chiral-even, leading-twist quark GPDs $H^q$ and $E^q$ of the proton. The construction combines a low-scale longitudinal input with flavor-dependent transverse profiles whose parameters are calibrated simultaneously against phenomenological representations of the proton and neutron electromagnetic form factors. In this way, the same set of quark distributions describes the flavor-separated Dirac and Pauli sectors while preserving the required charge and anomalous-magnetic-moment normalizations.

The simultaneous treatment of the four nucleon form factors is an important element of the analysis. Although the proton observables are dominated by the $u$-quark contribution, the neutron form factors provide essential information for separating the $u$- and $d$-quark profiles. The optimized distributions provide a reasonable simultaneous description of the target electromagnetic parametrization over the calibrated interval $0.01\leq Q^2\leq4.0~\mathrm{GeV}^2$. Nevertheless, this agreement must be interpreted with appropriate care. The calibration is performed against a phenomenological representation of the elastic data rather than directly against the complete set of experimental measurements and their correlations. Consequently, the resulting sensitivity bands reflect the adopted calibration tolerances and the chosen ansatz; they should not be interpreted as statistical confidence intervals or as the uncertainties of a global experimental likelihood analysis.

A central outcome is that the helicity-flip distribution cannot, in general, be assumed to have a forward shape proportional to the corresponding valence PDF. Although such a choice preserves the Pauli-form-factor normalization, it fixes the second moment too rigidly and may be incompatible with the angular-momentum and transverse-density constraints. We therefore employ an independently parametrized, normalization-preserving forward limit for $E^q$, with sufficient shape flexibility to modify its $x$-weighted moment without altering $F_2^q(0)$. The flavor parameters are related so that
\begin{equation}
B^{u+d}(0,\mu_0)=0
\end{equation}
is satisfied within numerical precision in the adopted quark-only low-scale description. Ji's relation then gives
\begin{align}
J^{u+d}(\mu_0) &= \frac{1}{2} \left[ A^{u+d}(0,\mu_0)+B^{u+d}(0,\mu_0) \right] \nonumber \\
&
= \frac{1}{2}A^{u+d}(0,\mu_0).
\end{align}
For the final parametrization, $A^{u+d}(0,\mu_0)=1.0005$ and $J^{u+d}(\mu_0)=0.5003$. Thus, the modeled quark sector saturates the proton spin to the accuracy of the adopted low-scale momentum matching. It is important, however, to keep the two conditions conceptually separate: $B^{u+d}(0,\mu_0)=0$ is imposed through the parametrization of $E^q$, whereas $A^{u+d}(0,\mu_0)\simeq1$ follows from the chosen low-scale valence input. Neither result follows from electromagnetic normalization alone.

The impact-parameter representation provides a further and independent test of the reconstructed distributions. For the Gaussian ansatz used here, positivity requires more than positive profile functions: the helicity-flip distortion generated by $E^q$ must remain bounded by the unpolarized density generated by $H^q$. The optimized profiles satisfy the required ordering $0<g_q(x)<f_q(x)$, and the complete transverse-density inequality specialized to the present Gaussian ansatz provides the pointwise criterion used to validate the parametrization. This analysis illustrates why reproducing electromagnetic form factors is not, by itself, sufficient to establish consistency with the zero-skewness transverse-density positivity condition. Form-factor sum rules constrain integrals over $x$, whereas positivity probes the local interplay among the longitudinal input, the Pauli distribution, and the transverse profiles.

We extended the zero-skewness distributions to $\xi\neq0$ through normalized double distributions. The construction has the correct spectral support, preserves the electromagnetic first moments, and generates the even powers of $\xi$ required by time-reversal invariance and polynomiality. The explicit reduction of the double-distribution integrals also provides useful analytic and numerical checks of the implementation. At the same time, the analysis makes clear that the double-distribution profile is an additional model input. Its width is not constrained by elastic form factors, even when the zero-skewness distributions have been determined accurately. Moreover, although the double-distribution representation guarantees the required support and polynomial structure, it does not automatically ensure the general positivity inequalities at nonzero skewness. Establishing those constraints would require a separate analysis beyond the zero-skewness transverse-density test performed here.

The highest power allowed in the relevant Mellin polynomials is supplied by an independent $D$-term. Because this contribution vanishes from the electromagnetic sum rules, neither its magnitude nor its momentum-transfer dependence can be extracted from the elastic calibration performed here. The $D$-term must therefore be regarded as a genuinely independent part of the mechanical sector. Higher Gegenbauer harmonics may change the distribution in the ERBL region without modifying the lowest energy--momentum-tensor moment, which further illustrates the nonuniqueness of the reconstruction. The same $D$-term enters $H^q$ and $E^q$ with opposite signs and therefore cancels from the angular-momentum combination $H^q+E^q$, as required.

As an analytic illustration of the mechanical construction, we considered a tripole parametrization of an independently specified total $D$-form factor. Its three-dimensional Fourier transform yields finite isotropic-pressure and shear-force distributions, satisfies the von Laue equilibrium condition, and leads to a positive radial pressure for $D_0<0$. These results characterize the chosen benchmark and should not be interpreted as an extraction of the physical proton stress distributions from electromagnetic data.

The present reconstruction should not be interpreted as a unique determination of the proton GPDs. Electromagnetic form factors fix only their first moments, whereas the longitudinal input, factorization scale, functional form of $E^q$, transverse profiles, double-distribution width, and $D$-term introduce distinct sources of model dependence. Experimental uncertainty, parametrization uncertainty, and completion uncertainty should therefore be propagated and reported separately. A more comprehensive analysis would fit the underlying elastic data with their correlations, explore alternative longitudinal and transverse ansatz families, and implement perturbative evolution within a consistently matched quark--gluon description. Comparison with deeply virtual observables would then test information not already contained in the electromagnetic calibration.

The main result of this work is thus a controlled elastic baseline with explicit normalization, polynomiality, angular-momentum, endpoint, and zero-skewness transverse-density positivity checks. This baseline supplies the conventions and numerical infrastructure required for the next stage of our program: the construction of nucleon-to-resonance transition GPDs. Such an extension is not a direct replacement of elastic form factors by transition form factors. Unequal external masses, different spin-parity assignments, resonance widths, and the off-diagonal nature of the matrix element require a transition-specific Lorentz decomposition and a separate interpretation of the Mellin moments. The nucleon-to-Roper transition provides a natural first application. Once its electromagnetic and local-operator form-factor conventions have been fixed consistently, the methods developed here can be adapted to construct and test its nonforward distributions, and subsequently those of other prominent nucleon resonances.


\appendix

\section{Analytic moments of the forward distributions}
\label{app:moments}

The beta-function forms adopted for the forward distributions allow their moments and normalization conditions to be evaluated analytically. For the valence distribution in Eq.~\eqref{eq:valence-input}, one has
\begin{equation}
\int_0^1 \dd x\,x^n q_v(x) = n_q \frac{B(a_q+n+1,b_q+1)}{B(a_q+1,b_q+1)}.
\label{eq:appendix-valence-moment}
\end{equation}
In particular, defining the normalized distribution
\begin{equation}
\rho_q(x)=\frac{q_v(x)}{n_q},
\end{equation}
its first moment and variance are
\begin{align}
\langle x\rangle_{\rho_q}
&=
\frac{a_q+1}{a_q+b_q+2},
\label{eq:appendix-valence-mean}
\\
\operatorname{Var}_{\rho_q}(x)
&=
\frac{(a_q+1)(b_q+1)}
     {(a_q+b_q+2)^2(a_q+b_q+3)}.
\label{eq:appendix-valence-variance}
\end{align}
For the input parameters in Eq.~\eqref{eq:valence-exponents}, these expressions give
\begin{equation}
\langle x\rangle_{\rho_u}=0.344,
\qquad
\langle x\rangle_{\rho_d}=0.3125.
\end{equation}
Multiplication by the valence numbers gives
\begin{equation}
A^u(0)=n_u\langle x\rangle_{\rho_u},
\qquad
A^d(0)=n_d\langle x\rangle_{\rho_d},
\end{equation}
and hence the value of $A^{u+d}(0)$ quoted in Sec.~\ref{sec:gravitational}.

The moments of the forward limit of $E_v^q$ can be treated in the same way. For compactness, define
\begin{equation}
\mathcal B_{q,n} = B(\alpha_q+n+1,\beta_q+1).
\label{eq:appendix-beta-E}
\end{equation}
Equations~\eqref{eq:forward-E-flexible} and~\eqref{eq:E-normalization} then give
\begin{equation}
\int_0^1\dd x\,x^n e_v^q(x)
=
\kappa_q
\frac{\mathcal B_{q,n}
      +\gamma_q\mathcal B_{q,n+1}}
     {\mathcal B_{q,0}
      +\gamma_q\mathcal B_{q,1}}.
\label{eq:appendix-E-moment}
\end{equation}
For $n=0$, Eq.~\eqref{eq:appendix-E-moment} reduces identically to the anomalous-magnetic-moment normalization,
\begin{equation}
\int_0^1\dd x\,e_v^q(x)=\kappa_q.
\end{equation}
For $n=1$, it gives the anomalous gravitomagnetic form factor,
\begin{equation}
B^q(0)
=
\kappa_q
\frac{\mathcal B_{q,1}
      +\gamma_q\mathcal B_{q,2}}
     {\mathcal B_{q,0}
      +\gamma_q\mathcal B_{q,1}},
\label{eq:appendix-Bq-zero}
\end{equation}
in agreement with Eq.~\eqref{eq:Bq-zero}.

The condition $B^u(0)+B^d(0)=0$ can be used to eliminate one of the Pauli-shape parameters analytically. In the implementation adopted here, $\gamma_d$ is treated as the derived parameter. Defining
\begin{equation}
r_d=-\frac{B^u(0)}{\kappa_d},
\label{eq:appendix-rd}
\end{equation}
Eq.~\eqref{eq:appendix-Bq-zero} gives
\begin{equation}
\gamma_d
=
\frac{\mathcal B_{d,1}
      -r_d\mathcal B_{d,0}}
     {r_d\mathcal B_{d,1}
      -\mathcal B_{d,2}},
\label{eq:appendix-gamma-d}
\end{equation}
provided the denominator is nonzero and the resulting forward distribution satisfies the endpoint and positivity requirements. Equation~\eqref{eq:appendix-gamma-d} makes explicit that the normalization $\int\dd x\,e_v^d=\kappa_d$ and the second-moment condition $B^u(0)+B^d(0)=0$ constrain different combinations of the shape parameters.

The elastic reconstruction nevertheless remains nonunique. To see this, consider a smooth function $\phi(x,t)$ that vanishes at $x=0$ and $x=1$ and satisfies $\phi(x,0)=0$. The modification
\begin{equation}
\delta H(x,0,t) = \frac{\partial\phi(x,t)}{\partial x}
\label{eq:appendix-null-direction}
\end{equation}
obeys
\begin{equation}
\int_0^1\dd x\,\delta H(x,0,t)=0,
\qquad
\delta H(x,0,0)=0.
\end{equation}
It therefore leaves both the elastic first moment and the forward PDF unchanged, while generally modifying the $x$ dependence at nonzero $t$ and all higher Mellin moments. Such a modification need not satisfy the positivity or endpoint constraints, but it illustrates explicitly the information that is not fixed by the form-factor sum rule alone.


\section{Moments of the double-distribution representation}
\label{app:DD-moments}

The moments of the profile function in the double-distribution ansatz can also be evaluated analytically. At fixed $\beta$, introduce
\begin{equation}
z=\frac{\alpha}{1-\beta}.
\end{equation}
The profile measure then becomes
\begin{equation}
h_b(\beta,\alpha)\dd\alpha
=
\frac{\Gamma(b+3/2)}
     {\sqrt{\pi}\,\Gamma(b+1)}
(1-z^2)^b\dd z.
\label{eq:appendix-DD-profile-measure}
\end{equation}
Odd moments in $\alpha$ vanish, whereas the even moments are
\begin{equation}
\left\langle\alpha^{2k}\right\rangle_\beta
=
(1-\beta)^{2k}
\frac{\Gamma(k+1/2)\Gamma(b+3/2)}
     {\sqrt{\pi}\,\Gamma(b+k+3/2)}.
\label{eq:appendix-alpha-moment}
\end{equation}
For $k=0$, this relation gives unity and thereby verifies the normalization of the profile. For $k=1$, it gives
\begin{equation}
\left\langle\alpha^2\right\rangle_\beta = \frac{(1-\beta)^2}{2b+3}.
\label{eq:appendix-alpha-second-moment}
\end{equation}

Combining Eq.~\eqref{eq:appendix-alpha-moment} with the binomial expansion provides a one-dimensional representation of every DD Mellin moment:
\begin{align}
&\int_{-1}^{1}\dd x\,x^N H_{\rm DD}^q(x,\xi,t)
\nonumber\\
&\quad =
\sum_{k=0}^{\lfloor N/2\rfloor}
\binom{N}{2k}\xi^{2k}
\frac{\Gamma(k+1/2)\Gamma(b+3/2)}
     {\sqrt{\pi}\,\Gamma(b+k+3/2)}
\nonumber\\
&\qquad\quad\times
\int_0^1\dd\beta\,
\beta^{N-2k}(1-\beta)^{2k}
q_v(\beta)e^{t f_q(\beta)}.
\label{eq:appendix-general-DD-moment}
\end{align}
This expression makes the even polynomial dependence on $\xi$ manifest.
The corresponding result for $E_{\rm DD}^q$ is obtained through the
replacements
\begin{equation}
q_v(\beta)\longrightarrow e_v^q(\beta),
\qquad
f_q(\beta)\longrightarrow g_q(\beta).
\end{equation}

For the first three moments of the complete GPD $H^q$, including its independent $D$-term contribution, one obtains
\begin{align}
\int_{-1}^{1}\dd x\,H^q(x,\xi,t)
&=
F_1^q(t),
\label{eq:appendix-H-first-moment}
\\
\int_{-1}^{1}\dd x\,xH^q(x,\xi,t)
&=
A^q(t)+\xi^2D^q(t),
\label{eq:appendix-H-second-moment}
\\
\int_{-1}^{1}\dd x\,x^2H^q(x,\xi,t)
&=
\int_0^1\dd\beta\,
q_v(\beta)e^{t f_q(\beta)}
\nonumber\\
&\quad\times
\left[
\beta^2+
\frac{\xi^2(1-\beta)^2}{2b+3}
\right].
\label{eq:appendix-H-third-moment}
\end{align}
The $D$-term does not contribute to Eq.~\eqref{eq:appendix-H-third-moment} because $\mathcal D^q(z,t)$ is odd in $z$. The analogous expressions for $E^q$ contain the forward input $e_v^q(\beta)e^{t g_q(\beta)}$ and the opposite sign for the $D$-term contribution to the second moment:
\begin{equation}
\int_{-1}^{1}\dd x\,xE^q(x,\xi,t)
=
B^q(t)-\xi^2D^q(t).
\label{eq:appendix-E-second-moment}
\end{equation}
Higher Mellin moments probe both the width of the double-distribution profile and the Gegenbauer coefficients of the $D$-term that are invisible to the spin-two form factors.


\section{Maximization of the transverse-density ratio}
\label{app:positivity}

At fixed $x$, let
\begin{equation}
f=f_q(x),
\qquad
g=g_q(x),
\qquad
b=b_\perp.
\end{equation}
For $0<g<f$, the Gaussian ratio in Eq.~\eqref{eq:Gaussian-positivity-ratio} can be written as
\begin{equation}
R_q(x,b)=K_q(x)\,b\,e^{-c_q(x)b^2},
\label{eq:appendix-R-form}
\end{equation}
where
\begin{equation}
K_q(x) = \frac{|e_v^q(x)|f_q(x)}{4M g_q^2(x)q_v(x)}, \qquad c_q(x) = \frac{f_q(x)-g_q(x)}{4f_q(x)g_q(x)} > 0.
\label{eq:appendix-K-c}
\end{equation}
Differentiation with respect to $b$ gives
\begin{equation}
\frac{\partial R_q}{\partial b} = K_q e^{-c_qb^2} \left(1-2c_qb^2\right).
\end{equation}
The maximum therefore occurs at
\begin{equation}
b_{\perp,\mathrm{max}}^2(x) = \frac{1}{2c_q(x)} = \frac{2f_q(x)g_q(x)}{f_q(x)-g_q(x)},
\end{equation}
in agreement with Eq.~\eqref{eq:bmax-positivity}. Substitution into Eq.~\eqref{eq:appendix-R-form} gives
\begin{equation}
R_{q,\mathrm{max}}^2(x) = \frac{[e_v^q(x)]^2}{[q_v(x)]^2} \frac{f_q^3(x)}{8eM^2g_q^3(x)[f_q(x)-g_q(x)]}.
\label{eq:appendix-Rmax}
\end{equation}
Requiring $R_{q,\mathrm{max}}\leq1$ yields
\begin{equation}
[e_v^q(x)]^2 \leq 8eM^2 \left[\frac{g_q(x)}{f_q(x)}\right]^3 [f_q(x)-g_q(x)] [q_v(x)]^2,
\end{equation}
which is Eq.~\eqref{eq:complete-Gaussian-bound}.

The derivation also explains why positivity of the two profile functions is not sufficient. If $g_q(x)>f_q(x)$, the exponential factor in Eq.~\eqref{eq:Gaussian-positivity-ratio} grows without bound at large $b_\perp$. If $g_q(x)=f_q(x)$, the ratio grows linearly with $b_\perp$. Consequently, the strict ordering $0<g_q(x)<f_q(x)$ is necessary wherever $e_v^q(x)\neq0$, and the complete bound above must then be verified pointwise in $x$.


\section{Fourier-transform and stress-distribution checks}
\label{app:stress-checks}

A convenient starting point for evaluating the three-dimensional transform of the tripole form factor is the Yukawa integral
\begin{equation}
\int\frac{\dd^3\boldsymbol k}{(2\pi)^3} \frac{e^{i\boldsymbol k\cdot\boldsymbol r}} {\boldsymbol k^2+m^2} = \frac{e^{-mr}}{4\pi r}.
\label{eq:appendix-Yukawa-transform}
\end{equation}
The identity
\begin{equation}
\frac{1}{(\boldsymbol k^2+m^2)^3} = \frac{1}{2} \frac{\partial^2}{\partial(m^2)^2} \frac{1}{\boldsymbol k^2+m^2}
\label{eq:appendix-cubic-identity}
\end{equation}
then gives
\begin{equation}
\int\frac{\dd^3\boldsymbol k}{(2\pi)^3}
\frac{e^{i\boldsymbol k\cdot\boldsymbol r}} {(\boldsymbol k^2+m^2)^3} = \frac{e^{-mr}}{32\pi m^3}(1+mr).
\label{eq:appendix-cubic-transform}
\end{equation}
Multiplication by $D_0m_D^6$ reproduces Eq.~\eqref{eq:D-tripole-transform}:
\begin{equation}
\widetilde D(r) = \frac{D_0m_D^3}{32\pi}(1+m_Dr)e^{-m_Dr}.
\end{equation}
This analytic expression avoids the need for a cutoff-dependent numerical Fourier transform.

Writing
\begin{equation}
K=\frac{D_0m_D^3}{32\pi},
\end{equation}
the derivatives required in Eqs.~\eqref{eq:pressure-from-D} and~\eqref{eq:shear-from-D} are
\begin{align}
\widetilde D'(r)
&=
-Km_D^2r e^{-m_Dr},
\\
\widetilde D''(r)
&=
Km_D^2(m_Dr-1)e^{-m_Dr}.
\end{align}
Their substitution reproduces the pressure, shear-force, and radial-pressure distributions in Eqs.~\eqref{eq:tripole-pressure}--\eqref{eq:tripole-radial-pressure}.

The von Laue condition follows directly from the pressure definition:
\begin{align}
\int_0^\infty\dd r\,r^2p(r)
&=
\frac{1}{6M} \int_0^\infty\dd r\, \frac{\dd}{\dd r} \left[r^2\widetilde D'(r)\right] \nonumber\\
&=
\frac{1}{6M} \left[ r^2\widetilde D'(r) \right]_{0}^{\infty} =0,
\label{eq:appendix-von-Laue-proof}
\end{align}
because the surface term vanishes at both endpoints. This proves Eq.~\eqref{eq:von-Laue} for the tripole benchmark.

A further normalization check is
\begin{equation}
M\int\dd^3\boldsymbol r\,r^2p(r)=D_0.
\label{eq:appendix-pressure-normalization}
\end{equation}
Using Eq.~\eqref{eq:tripole-pressure}, this relation reduces to
\begin{equation}
\frac{D_0}{48} \int_0^\infty\dd y\, y^4(y-3)e^{-y} = D_0,
\end{equation}
since
\begin{equation}
\int_0^\infty\dd y\, y^4(y-3)e^{-y} = 5!-3\,4! = 48.
\end{equation}
All these relations refer to the static Breit-frame convention specified in Sec.~\ref{sec:mechanical-distributions} and to the illustrative form factor of the complete conserved energy--momentum tensor.


\begin{acknowledgments}
J.S. is grateful to J. M. Morgado-Chávez for stimulating and insightful discussions.
Work partially financed by 
the Escuela Polit\'ecnica Nacional under projects PIS-22-04 and PIM 23-01; 
HADRON-2030 which receives funding from the European Union's Horizon Europe research and innovation programme under Grant Agreement No. 101292522;
Ministerio Espa\~nol de Ciencia e Innovaci\'on under grant nos. PID2022-141910NB-I00 and PID2022-140440NB-C22;
Junta de Castilla y Le\'on grant SA091P24 under program EDU/841/2024;
Junta de Andaluc\'ia under contract no. PAIDI FQM-370 and and PCI+D+i under the title: ``Tecnologías avanzadas para la exploración del universo y sus componentes” (Code AST22-0001).
During the preparation of this manuscript, the authors used OpenAI ChatGPT (GPT-5) as an interactive assistant for organizing and revising portions of the text, improving the pedagogical presentation, and checking notation and internal consistency. The authors directed the tool through targeted questions based on their own calculations and manuscript, independently verified all retained suggestions, and critically reviewed and edited the resulting text. All numerical calculations, parameter calibrations, data analysis, and figures were produced and validated by the authors, who take full responsibility for the content of the manuscript.
\end{acknowledgments}


\bibliography{print_GPDsPhenoNucleon}

\begin{thebibliography}{28}%
\makeatletter
\providecommand \@ifxundefined [1]{%
 \@ifx{#1\undefined}
}%
\providecommand \@ifnum [1]{%
 \ifnum #1\expandafter \@firstoftwo
 \else \expandafter \@secondoftwo
 \fi
}%
\providecommand \@ifx [1]{%
 \ifx #1\expandafter \@firstoftwo
 \else \expandafter \@secondoftwo
 \fi
}%
\providecommand \natexlab [1]{#1}%
\providecommand \enquote  [1]{``#1''}%
\providecommand \bibnamefont  [1]{#1}%
\providecommand \bibfnamefont [1]{#1}%
\providecommand \citenamefont [1]{#1}%
\providecommand \href@noop [0]{\@secondoftwo}%
\providecommand \href [0]{\begingroup \@sanitize@url \@href}%
\providecommand \@href[1]{\@@startlink{#1}\@@href}%
\providecommand \@@href[1]{\endgroup#1\@@endlink}%
\providecommand \@sanitize@url [0]{\catcode `\\12\catcode `\$12\catcode
  `\&12\catcode `\#12\catcode `\^12\catcode `\_12\catcode `\%12\relax}%
\providecommand \@@startlink[1]{}%
\providecommand \@@endlink[0]{}%
\providecommand \url  [0]{\begingroup\@sanitize@url \@url }%
\providecommand \@url [1]{\endgroup\@href {#1}{\urlprefix }}%
\providecommand \urlprefix  [0]{URL }%
\providecommand \Eprint [0]{\href }%
\providecommand \doibase [0]{https://doi.org/}%
\providecommand \selectlanguage [0]{\@gobble}%
\providecommand \bibinfo  [0]{\@secondoftwo}%
\providecommand \bibfield  [0]{\@secondoftwo}%
\providecommand \translation [1]{[#1]}%
\providecommand \BibitemOpen [0]{}%
\providecommand \bibitemStop [0]{}%
\providecommand \bibitemNoStop [0]{.\EOS\space}%
\providecommand \EOS [0]{\spacefactor3000\relax}%
\providecommand \BibitemShut  [1]{\csname bibitem#1\endcsname}%
\let\auto@bib@innerbib\@empty
\bibitem [{\citenamefont {M{\"u}ller}\ \emph {et~al.}(1994)\citenamefont
  {M{\"u}ller}, \citenamefont {Robaschik}, \citenamefont {Geyer}, \citenamefont
  {Dittes},\ and\ \citenamefont {Ho{\v{r}}ej{\v{s}}i}}]{Mueller1994}%
  \BibitemOpen
  \bibfield  {author} {\bibinfo {author} {\bibfnamefont {D.}~\bibnamefont
  {M{\"u}ller}}, \bibinfo {author} {\bibfnamefont {D.}~\bibnamefont
  {Robaschik}}, \bibinfo {author} {\bibfnamefont {B.}~\bibnamefont {Geyer}},
  \bibinfo {author} {\bibfnamefont {F.-M.}\ \bibnamefont {Dittes}},\ and\
  \bibinfo {author} {\bibfnamefont {J.}~\bibnamefont {Ho{\v{r}}ej{\v{s}}i}},\
  }\bibfield  {title} {\bibinfo {title} {Wave functions, evolution equations
  and evolution kernels from light-ray operators of {QCD}},\ }\href
  {https://doi.org/10.1002/prop.2190420202} {\bibfield  {journal} {\bibinfo
  {journal} {Fortschr. Phys.}\ }\textbf {\bibinfo {volume} {42}},\ \bibinfo
  {pages} {101} (\bibinfo {year} {1994})},\ \Eprint
  {https://arxiv.org/abs/hep-ph/9812448} {arXiv:hep-ph/9812448} \BibitemShut
  {NoStop}%
\bibitem [{\citenamefont {Ji}(1997{\natexlab{a}})}]{JiDVCS1997}%
  \BibitemOpen
  \bibfield  {author} {\bibinfo {author} {\bibfnamefont {X.-D.}\ \bibnamefont
  {Ji}},\ }\bibfield  {title} {\bibinfo {title} {Deeply virtual {Compton}
  scattering},\ }\href {https://doi.org/10.1103/PhysRevD.55.7114} {\bibfield
  {journal} {\bibinfo  {journal} {Phys. Rev. D}\ }\textbf {\bibinfo {volume}
  {55}},\ \bibinfo {pages} {7114} (\bibinfo {year} {1997}{\natexlab{a}})},\
  \Eprint {https://arxiv.org/abs/hep-ph/9609381} {arXiv:hep-ph/9609381}
  \BibitemShut {NoStop}%
\bibitem [{\citenamefont {Radyushkin}(1996)}]{Radyushkin1996}%
  \BibitemOpen
  \bibfield  {author} {\bibinfo {author} {\bibfnamefont {A.~V.}\ \bibnamefont
  {Radyushkin}},\ }\bibfield  {title} {\bibinfo {title} {Scaling limit of
  deeply virtual {Compton} scattering},\ }\href
  {https://doi.org/10.1016/0370-2693(96)00528-X} {\bibfield  {journal}
  {\bibinfo  {journal} {Phys. Lett. B}\ }\textbf {\bibinfo {volume} {380}},\
  \bibinfo {pages} {417} (\bibinfo {year} {1996})},\ \Eprint
  {https://arxiv.org/abs/hep-ph/9604317} {arXiv:hep-ph/9604317} \BibitemShut
  {NoStop}%
\bibitem [{\citenamefont {Diehl}(2003)}]{Diehl2003}%
  \BibitemOpen
  \bibfield  {author} {\bibinfo {author} {\bibfnamefont {M.}~\bibnamefont
  {Diehl}},\ }\bibfield  {title} {\bibinfo {title} {Generalized parton
  distributions},\ }\href {https://doi.org/10.1016/j.physrep.2003.08.002}
  {\bibfield  {journal} {\bibinfo  {journal} {Phys. Rept.}\ }\textbf {\bibinfo
  {volume} {388}},\ \bibinfo {pages} {41} (\bibinfo {year} {2003})},\ \Eprint
  {https://arxiv.org/abs/hep-ph/0307382} {arXiv:hep-ph/0307382 [hep-ph]}
  \BibitemShut {NoStop}%
\bibitem [{\citenamefont {Belitsky}\ and\ \citenamefont
  {Radyushkin}(2005)}]{Belitsky2005}%
  \BibitemOpen
  \bibfield  {author} {\bibinfo {author} {\bibfnamefont {A.~V.}\ \bibnamefont
  {Belitsky}}\ and\ \bibinfo {author} {\bibfnamefont {A.~V.}\ \bibnamefont
  {Radyushkin}},\ }\bibfield  {title} {\bibinfo {title} {Unraveling hadron
  structure with generalized parton distributions},\ }\href
  {https://doi.org/10.1016/j.physrep.2005.06.002} {\bibfield  {journal}
  {\bibinfo  {journal} {Phys. Rept.}\ }\textbf {\bibinfo {volume} {418}},\
  \bibinfo {pages} {1} (\bibinfo {year} {2005})},\ \Eprint
  {https://arxiv.org/abs/hep-ph/0504030} {arXiv:hep-ph/0504030 [hep-ph]}
  \BibitemShut {NoStop}%
\bibitem [{\citenamefont {Collins}\ \emph {et~al.}(1997)\citenamefont
  {Collins}, \citenamefont {Frankfurt},\ and\ \citenamefont
  {Strikman}}]{CollinsFrankfurtStrikman1997}%
  \BibitemOpen
  \bibfield  {author} {\bibinfo {author} {\bibfnamefont {J.~C.}\ \bibnamefont
  {Collins}}, \bibinfo {author} {\bibfnamefont {L.}~\bibnamefont {Frankfurt}},\
  and\ \bibinfo {author} {\bibfnamefont {M.}~\bibnamefont {Strikman}},\
  }\bibfield  {title} {\bibinfo {title} {Factorization for hard exclusive
  electroproduction of mesons in {QCD}},\ }\href
  {https://doi.org/10.1103/PhysRevD.56.2982} {\bibfield  {journal} {\bibinfo
  {journal} {Phys. Rev. D}\ }\textbf {\bibinfo {volume} {56}},\ \bibinfo
  {pages} {2982} (\bibinfo {year} {1997})},\ \Eprint
  {https://arxiv.org/abs/hep-ph/9611433} {arXiv:hep-ph/9611433} \BibitemShut
  {NoStop}%
\bibitem [{\citenamefont {Belitsky}\ \emph {et~al.}(2002)\citenamefont
  {Belitsky}, \citenamefont {M{\"u}ller},\ and\ \citenamefont
  {Kirchner}}]{BelitskyMullerKirchner2002}%
  \BibitemOpen
  \bibfield  {author} {\bibinfo {author} {\bibfnamefont {A.~V.}\ \bibnamefont
  {Belitsky}}, \bibinfo {author} {\bibfnamefont {D.}~\bibnamefont
  {M{\"u}ller}},\ and\ \bibinfo {author} {\bibfnamefont {A.}~\bibnamefont
  {Kirchner}},\ }\bibfield  {title} {\bibinfo {title} {Theory of deeply virtual
  {Compton} scattering on the nucleon},\ }\href
  {https://doi.org/10.1016/S0550-3213(02)00144-X} {\bibfield  {journal}
  {\bibinfo  {journal} {Nucl. Phys. B}\ }\textbf {\bibinfo {volume} {629}},\
  \bibinfo {pages} {323} (\bibinfo {year} {2002})},\ \Eprint
  {https://arxiv.org/abs/hep-ph/0112108} {arXiv:hep-ph/0112108} \BibitemShut
  {NoStop}%
\bibitem [{\citenamefont {Abdul~Khalek}\ \emph {et~al.}(2022)\citenamefont
  {Abdul~Khalek} \emph {et~al.}}]{EICYellowReport2022}%
  \BibitemOpen
  \bibfield  {author} {\bibinfo {author} {\bibfnamefont {R.}~\bibnamefont
  {Abdul~Khalek}} \emph {et~al.} (\bibinfo {collaboration} {EIC Yellow
  Report}),\ }\bibfield  {title} {\bibinfo {title} {Science requirements and
  detector concepts for the {Electron-Ion Collider}: {EIC Yellow Report}},\
  }\href {https://doi.org/10.1016/j.nuclphysa.2022.122447} {\bibfield
  {journal} {\bibinfo  {journal} {Nucl. Phys. A}\ }\textbf {\bibinfo {volume}
  {1026}},\ \bibinfo {pages} {122447} (\bibinfo {year} {2022})},\ \Eprint
  {https://arxiv.org/abs/2103.05419} {arXiv:2103.05419 [physics.ins-det]}
  \BibitemShut {NoStop}%
\bibitem [{\citenamefont {Ji}(1997{\natexlab{b}})}]{Ji1997}%
  \BibitemOpen
  \bibfield  {author} {\bibinfo {author} {\bibfnamefont {X.-D.}\ \bibnamefont
  {Ji}},\ }\bibfield  {title} {\bibinfo {title} {Gauge-invariant decomposition
  of nucleon spin},\ }\href {https://doi.org/10.1103/PhysRevLett.78.610}
  {\bibfield  {journal} {\bibinfo  {journal} {Phys. Rev. Lett.}\ }\textbf
  {\bibinfo {volume} {78}},\ \bibinfo {pages} {610} (\bibinfo {year}
  {1997}{\natexlab{b}})},\ \Eprint {https://arxiv.org/abs/hep-ph/9603249}
  {arXiv:hep-ph/9603249 [hep-ph]} \BibitemShut {NoStop}%
\bibitem [{\citenamefont {Burkardt}(2000)}]{Burkardt2000}%
  \BibitemOpen
  \bibfield  {author} {\bibinfo {author} {\bibfnamefont {M.}~\bibnamefont
  {Burkardt}},\ }\bibfield  {title} {\bibinfo {title} {Impact parameter
  dependent parton distributions and off-forward parton distributions for
  $\zeta\to 0$},\ }\href {https://doi.org/10.1103/PhysRevD.62.071503}
  {\bibfield  {journal} {\bibinfo  {journal} {Phys. Rev. D}\ }\textbf {\bibinfo
  {volume} {62}},\ \bibinfo {pages} {071503} (\bibinfo {year} {2000})},\
  \Eprint {https://arxiv.org/abs/hep-ph/0005108} {arXiv:hep-ph/0005108
  [hep-ph]} \BibitemShut {NoStop}%
\bibitem [{\citenamefont {Diehl}(2002)}]{Diehl2002}%
  \BibitemOpen
  \bibfield  {author} {\bibinfo {author} {\bibfnamefont {M.}~\bibnamefont
  {Diehl}},\ }\bibfield  {title} {\bibinfo {title} {Generalized parton
  distributions in impact parameter space},\ }\href
  {https://doi.org/10.1007/s10052-002-1016-9} {\bibfield  {journal} {\bibinfo
  {journal} {Eur. Phys. J. C}\ }\textbf {\bibinfo {volume} {25}},\ \bibinfo
  {pages} {223} (\bibinfo {year} {2002})},\ \Eprint
  {https://arxiv.org/abs/hep-ph/0205208} {arXiv:hep-ph/0205208 [hep-ph]}
  \BibitemShut {NoStop}%
\bibitem [{\citenamefont {Polyakov}\ and\ \citenamefont
  {Schweitzer}(2018)}]{PolyakovSchweitzer2018}%
  \BibitemOpen
  \bibfield  {author} {\bibinfo {author} {\bibfnamefont {M.~V.}\ \bibnamefont
  {Polyakov}}\ and\ \bibinfo {author} {\bibfnamefont {P.}~\bibnamefont
  {Schweitzer}},\ }\bibfield  {title} {\bibinfo {title} {Forces inside hadrons:
  Pressure, surface tension, mechanical radius, and all that},\ }\href
  {https://doi.org/10.1142/S0217751X18300259} {\bibfield  {journal} {\bibinfo
  {journal} {Int. J. Mod. Phys. A}\ }\textbf {\bibinfo {volume} {33}},\
  \bibinfo {pages} {1830025} (\bibinfo {year} {2018})},\ \Eprint
  {https://arxiv.org/abs/1805.06596} {arXiv:1805.06596 [hep-ph]} \BibitemShut
  {NoStop}%
\bibitem [{\citenamefont {Burkert}\ \emph {et~al.}(2018)\citenamefont
  {Burkert}, \citenamefont {Elouadrhiri},\ and\ \citenamefont
  {Girod}}]{Burkert2018}%
  \BibitemOpen
  \bibfield  {author} {\bibinfo {author} {\bibfnamefont {V.~D.}\ \bibnamefont
  {Burkert}}, \bibinfo {author} {\bibfnamefont {L.}~\bibnamefont
  {Elouadrhiri}},\ and\ \bibinfo {author} {\bibfnamefont {F.~X.}\ \bibnamefont
  {Girod}},\ }\bibfield  {title} {\bibinfo {title} {The pressure distribution
  inside the proton},\ }\href {https://doi.org/10.1038/s41586-018-0060-z}
  {\bibfield  {journal} {\bibinfo  {journal} {Nature}\ }\textbf {\bibinfo
  {volume} {557}},\ \bibinfo {pages} {396} (\bibinfo {year}
  {2018})}\BibitemShut {NoStop}%
\bibitem [{\citenamefont {Diehl}\ \emph {et~al.}(2005)\citenamefont {Diehl},
  \citenamefont {Feldmann}, \citenamefont {Jakob},\ and\ \citenamefont
  {Kroll}}]{Diehl2005}%
  \BibitemOpen
  \bibfield  {author} {\bibinfo {author} {\bibfnamefont {M.}~\bibnamefont
  {Diehl}}, \bibinfo {author} {\bibfnamefont {T.}~\bibnamefont {Feldmann}},
  \bibinfo {author} {\bibfnamefont {R.}~\bibnamefont {Jakob}},\ and\ \bibinfo
  {author} {\bibfnamefont {P.}~\bibnamefont {Kroll}},\ }\bibfield  {title}
  {\bibinfo {title} {Generalized parton distributions from nucleon form-factor
  data},\ }\href {https://doi.org/10.1140/epjc/s2004-02063-4} {\bibfield
  {journal} {\bibinfo  {journal} {Eur. Phys. J. C}\ }\textbf {\bibinfo {volume}
  {39}},\ \bibinfo {pages} {1} (\bibinfo {year} {2005})},\ \Eprint
  {https://arxiv.org/abs/hep-ph/0408173} {arXiv:hep-ph/0408173 [hep-ph]}
  \BibitemShut {NoStop}%
\bibitem [{\citenamefont {Diehl}\ and\ \citenamefont
  {Kroll}(2013)}]{DiehlKroll2013}%
  \BibitemOpen
  \bibfield  {author} {\bibinfo {author} {\bibfnamefont {M.}~\bibnamefont
  {Diehl}}\ and\ \bibinfo {author} {\bibfnamefont {P.}~\bibnamefont {Kroll}},\
  }\bibfield  {title} {\bibinfo {title} {Nucleon form factors, generalized
  parton distributions and quark angular momentum},\ }\href
  {https://doi.org/10.1140/epjc/s10052-013-2397-7} {\bibfield  {journal}
  {\bibinfo  {journal} {Eur. Phys. J. C}\ }\textbf {\bibinfo {volume} {73}},\
  \bibinfo {pages} {2397} (\bibinfo {year} {2013})},\ \Eprint
  {https://arxiv.org/abs/1302.4604} {arXiv:1302.4604 [hep-ph]} \BibitemShut
  {NoStop}%
\bibitem [{\citenamefont {Hashamipour}\ \emph {et~al.}(2023)\citenamefont
  {Hashamipour}, \citenamefont {Goharipour}, \citenamefont {Azizi},\ and\
  \citenamefont {Goloskokov}}]{Hashamipour2023}%
  \BibitemOpen
  \bibfield  {author} {\bibinfo {author} {\bibfnamefont {H.}~\bibnamefont
  {Hashamipour}}, \bibinfo {author} {\bibfnamefont {M.}~\bibnamefont
  {Goharipour}}, \bibinfo {author} {\bibfnamefont {K.}~\bibnamefont {Azizi}},\
  and\ \bibinfo {author} {\bibfnamefont {S.~V.}\ \bibnamefont {Goloskokov}},\
  }\bibfield  {title} {\bibinfo {title} {Generalized parton distributions at
  zero skewness},\ }\href {https://doi.org/10.1103/PhysRevD.107.096005}
  {\bibfield  {journal} {\bibinfo  {journal} {Phys. Rev. D}\ }\textbf {\bibinfo
  {volume} {107}},\ \bibinfo {pages} {096005} (\bibinfo {year} {2023})},\
  \Eprint {https://arxiv.org/abs/2211.09522} {arXiv:2211.09522 [hep-ph]}
  \BibitemShut {NoStop}%
\bibitem [{\citenamefont {Guidal}\ \emph {et~al.}(2013)\citenamefont {Guidal},
  \citenamefont {Moutarde},\ and\ \citenamefont {Vanderhaeghen}}]{Guidal2013}%
  \BibitemOpen
  \bibfield  {author} {\bibinfo {author} {\bibfnamefont {M.}~\bibnamefont
  {Guidal}}, \bibinfo {author} {\bibfnamefont {H.}~\bibnamefont {Moutarde}},\
  and\ \bibinfo {author} {\bibfnamefont {M.}~\bibnamefont {Vanderhaeghen}},\
  }\bibfield  {title} {\bibinfo {title} {Generalized parton distributions in
  the valence region from deeply virtual {Compton} scattering kinematics},\
  }\href {https://doi.org/10.1088/0034-4885/76/6/066202} {\bibfield  {journal}
  {\bibinfo  {journal} {Rept. Prog. Phys.}\ }\textbf {\bibinfo {volume} {76}},\
  \bibinfo {pages} {066202} (\bibinfo {year} {2013})},\ \Eprint
  {https://arxiv.org/abs/1303.6600} {arXiv:1303.6600 [hep-ph]} \BibitemShut
  {NoStop}%
\bibitem [{\citenamefont {Radyushkin}(1999)}]{Radyushkin1999}%
  \BibitemOpen
  \bibfield  {author} {\bibinfo {author} {\bibfnamefont {A.~V.}\ \bibnamefont
  {Radyushkin}},\ }\bibfield  {title} {\bibinfo {title} {Double distributions
  and evolution equations},\ }\href
  {https://doi.org/10.1103/PhysRevD.59.014030} {\bibfield  {journal} {\bibinfo
  {journal} {Phys. Rev. D}\ }\textbf {\bibinfo {volume} {59}},\ \bibinfo
  {pages} {014030} (\bibinfo {year} {1999})},\ \Eprint
  {https://arxiv.org/abs/hep-ph/9805342} {arXiv:hep-ph/9805342 [hep-ph]}
  \BibitemShut {NoStop}%
\bibitem [{\citenamefont {Polyakov}\ and\ \citenamefont
  {Weiss}(1999)}]{PolyakovWeiss1999}%
  \BibitemOpen
  \bibfield  {author} {\bibinfo {author} {\bibfnamefont {M.~V.}\ \bibnamefont
  {Polyakov}}\ and\ \bibinfo {author} {\bibfnamefont {C.}~\bibnamefont
  {Weiss}},\ }\bibfield  {title} {\bibinfo {title} {Skewed and double
  distributions in pion and nucleon},\ }\href
  {https://doi.org/10.1103/PhysRevD.60.114017} {\bibfield  {journal} {\bibinfo
  {journal} {Phys. Rev. D}\ }\textbf {\bibinfo {volume} {60}},\ \bibinfo
  {pages} {114017} (\bibinfo {year} {1999})},\ \Eprint
  {https://arxiv.org/abs/hep-ph/9902451} {arXiv:hep-ph/9902451 [hep-ph]}
  \BibitemShut {NoStop}%
\bibitem [{\citenamefont {Diehl}\ \emph {et~al.}(2025)\citenamefont {Diehl},
  \citenamefont {Joo}, \citenamefont {Semenov-Tian-Shansky}, \citenamefont
  {Weiss}, \citenamefont {Braun}, \citenamefont {Chang}, \citenamefont
  {Chatagnon}, \citenamefont {Constantinou}, \citenamefont {Guo}, \citenamefont
  {Hutauruk}, \citenamefont {Jo}, \citenamefont {Kim}, \citenamefont {Kim},
  \citenamefont {Kroll}, \citenamefont {Kumano}, \citenamefont {Lee},
  \citenamefont {Liuti}, \citenamefont {McNulty}, \citenamefont {Son},
  \citenamefont {Sznajder}, \citenamefont {Usman}, \citenamefont {Van~Hulse},
  \citenamefont {Vanderhaeghen},\ and\ \citenamefont {Winn}}]{Transition2025}%
  \BibitemOpen
  \bibfield  {author} {\bibinfo {author} {\bibfnamefont {S.}~\bibnamefont
  {Diehl}}, \bibinfo {author} {\bibfnamefont {K.}~\bibnamefont {Joo}}, \bibinfo
  {author} {\bibfnamefont {K.}~\bibnamefont {Semenov-Tian-Shansky}}, \bibinfo
  {author} {\bibfnamefont {C.}~\bibnamefont {Weiss}}, \bibinfo {author}
  {\bibfnamefont {V.}~\bibnamefont {Braun}}, \bibinfo {author} {\bibfnamefont
  {W.-C.}\ \bibnamefont {Chang}}, \bibinfo {author} {\bibfnamefont
  {P.}~\bibnamefont {Chatagnon}}, \bibinfo {author} {\bibfnamefont
  {M.}~\bibnamefont {Constantinou}}, \bibinfo {author} {\bibfnamefont
  {Y.}~\bibnamefont {Guo}}, \bibinfo {author} {\bibfnamefont {P.~T.~P.}\
  \bibnamefont {Hutauruk}}, \bibinfo {author} {\bibfnamefont {H.-S.}\
  \bibnamefont {Jo}}, \bibinfo {author} {\bibfnamefont {A.}~\bibnamefont
  {Kim}}, \bibinfo {author} {\bibfnamefont {J.-Y.}\ \bibnamefont {Kim}},
  \bibinfo {author} {\bibfnamefont {P.}~\bibnamefont {Kroll}}, \bibinfo
  {author} {\bibfnamefont {S.}~\bibnamefont {Kumano}}, \bibinfo {author}
  {\bibfnamefont {C.-H.}\ \bibnamefont {Lee}}, \bibinfo {author} {\bibfnamefont
  {S.}~\bibnamefont {Liuti}}, \bibinfo {author} {\bibfnamefont
  {R.}~\bibnamefont {McNulty}}, \bibinfo {author} {\bibfnamefont {H.-D.}\
  \bibnamefont {Son}}, \bibinfo {author} {\bibfnamefont {P.}~\bibnamefont
  {Sznajder}}, \bibinfo {author} {\bibfnamefont {A.}~\bibnamefont {Usman}},
  \bibinfo {author} {\bibfnamefont {C.}~\bibnamefont {Van~Hulse}}, \bibinfo
  {author} {\bibfnamefont {M.}~\bibnamefont {Vanderhaeghen}},\ and\ \bibinfo
  {author} {\bibfnamefont {M.}~\bibnamefont {Winn}},\ }\bibfield  {title}
  {\bibinfo {title} {Exploring baryon resonances with transition generalized
  parton distributions: Status and perspectives},\ }\href
  {https://doi.org/10.1140/epja/s10050-025-01552-2} {\bibfield  {journal}
  {\bibinfo  {journal} {Eur. Phys. J. A}\ }\textbf {\bibinfo {volume} {61}},\
  \bibinfo {pages} {131} (\bibinfo {year} {2025})},\ \Eprint
  {https://arxiv.org/abs/2405.15386} {arXiv:2405.15386 [hep-ph]} \BibitemShut
  {NoStop}%
\bibitem [{\citenamefont {Chang}\ \emph {et~al.}(2022)\citenamefont {Chang},
  \citenamefont {Gao},\ and\ \citenamefont {Roberts}}]{Chang2022}%
  \BibitemOpen
  \bibfield  {author} {\bibinfo {author} {\bibfnamefont {L.}~\bibnamefont
  {Chang}}, \bibinfo {author} {\bibfnamefont {F.}~\bibnamefont {Gao}},\ and\
  \bibinfo {author} {\bibfnamefont {C.~D.}\ \bibnamefont {Roberts}},\
  }\bibfield  {title} {\bibinfo {title} {Parton distributions of light quarks
  and antiquarks in the proton},\ }\href
  {https://doi.org/10.1016/j.physletb.2022.137078} {\bibfield  {journal}
  {\bibinfo  {journal} {Phys. Lett. B}\ }\textbf {\bibinfo {volume} {829}},\
  \bibinfo {pages} {137078} (\bibinfo {year} {2022})},\ \Eprint
  {https://arxiv.org/abs/2201.07870} {arXiv:2201.07870 [hep-ph]} \BibitemShut
  {NoStop}%
\bibitem [{\citenamefont {Chouika}\ \emph {et~al.}(2018)\citenamefont
  {Chouika}, \citenamefont {Mezrag}, \citenamefont {Moutarde},\ and\
  \citenamefont {Rodr{\'i}guez-Quintero}}]{Chouika2018}%
  \BibitemOpen
  \bibfield  {author} {\bibinfo {author} {\bibfnamefont {N.}~\bibnamefont
  {Chouika}}, \bibinfo {author} {\bibfnamefont {C.}~\bibnamefont {Mezrag}},
  \bibinfo {author} {\bibfnamefont {H.}~\bibnamefont {Moutarde}},\ and\
  \bibinfo {author} {\bibfnamefont {J.}~\bibnamefont
  {Rodr{\'i}guez-Quintero}},\ }\bibfield  {title} {\bibinfo {title} {Covariant
  extension of the {GPD} overlap representation at low {Fock} states},\ }\href
  {https://doi.org/10.1016/j.physletb.2018.03.044} {\bibfield  {journal}
  {\bibinfo  {journal} {Phys. Lett. B}\ }\textbf {\bibinfo {volume} {780}},\
  \bibinfo {pages} {287} (\bibinfo {year} {2018})},\ \Eprint
  {https://arxiv.org/abs/1711.05108} {arXiv:1711.05108 [hep-ph]} \BibitemShut
  {NoStop}%
\bibitem [{\citenamefont {Kelly}(2004)}]{Kelly2004}%
  \BibitemOpen
  \bibfield  {author} {\bibinfo {author} {\bibfnamefont {J.~J.}\ \bibnamefont
  {Kelly}},\ }\bibfield  {title} {\bibinfo {title} {Simple parametrization of
  nucleon form factors},\ }\href {https://doi.org/10.1103/PhysRevC.70.068202}
  {\bibfield  {journal} {\bibinfo  {journal} {Phys. Rev. C}\ }\textbf {\bibinfo
  {volume} {70}},\ \bibinfo {pages} {068202} (\bibinfo {year}
  {2004})}\BibitemShut {NoStop}%
\bibitem [{\citenamefont {Pobylitsa}(2002)}]{Pobylitsa2002}%
  \BibitemOpen
  \bibfield  {author} {\bibinfo {author} {\bibfnamefont {P.~V.}\ \bibnamefont
  {Pobylitsa}},\ }\bibfield  {title} {\bibinfo {title} {Positivity bounds on
  generalized parton distributions in the impact parameter representation},\
  }\href {https://doi.org/10.1103/PhysRevD.66.094002} {\bibfield  {journal}
  {\bibinfo  {journal} {Phys. Rev. D}\ }\textbf {\bibinfo {volume} {66}},\
  \bibinfo {pages} {094002} (\bibinfo {year} {2002})},\ \Eprint
  {https://arxiv.org/abs/hep-ph/0204337} {arXiv:hep-ph/0204337} \BibitemShut
  {NoStop}%
\bibitem [{\citenamefont {Radyushkin}(1997)}]{Radyushkin1997}%
  \BibitemOpen
  \bibfield  {author} {\bibinfo {author} {\bibfnamefont {A.~V.}\ \bibnamefont
  {Radyushkin}},\ }\bibfield  {title} {\bibinfo {title} {Nonforward parton
  distributions},\ }\href {https://doi.org/10.1103/PhysRevD.56.5524} {\bibfield
   {journal} {\bibinfo  {journal} {Phys. Rev. D}\ }\textbf {\bibinfo {volume}
  {56}},\ \bibinfo {pages} {5524} (\bibinfo {year} {1997})},\ \Eprint
  {https://arxiv.org/abs/hep-ph/9704207} {arXiv:hep-ph/9704207 [hep-ph]}
  \BibitemShut {NoStop}%
\bibitem [{\citenamefont {Dutrieux}\ \emph {et~al.}(2025)\citenamefont
  {Dutrieux}, \citenamefont {Meisgny}, \citenamefont {Mezrag},\ and\
  \citenamefont {Moutarde}}]{Dutrieux2025}%
  \BibitemOpen
  \bibfield  {author} {\bibinfo {author} {\bibfnamefont {H.}~\bibnamefont
  {Dutrieux}}, \bibinfo {author} {\bibfnamefont {T.}~\bibnamefont {Meisgny}},
  \bibinfo {author} {\bibfnamefont {C.}~\bibnamefont {Mezrag}},\ and\ \bibinfo
  {author} {\bibfnamefont {H.}~\bibnamefont {Moutarde}},\ }\bibfield  {title}
  {\bibinfo {title} {Proton internal pressure from deeply virtual compton
  scattering on collider kinematics},\ }\href
  {https://doi.org/10.1140/epjc/s10052-024-13737-y} {\bibfield  {journal}
  {\bibinfo  {journal} {Eur. Phys. J. C}\ }\textbf {\bibinfo {volume} {85}},\
  \bibinfo {pages} {105} (\bibinfo {year} {2025})},\ \Eprint
  {https://arxiv.org/abs/2410.13518} {arXiv:2410.13518 [hep-ph]} \BibitemShut
  {NoStop}%
\bibitem [{\citenamefont {Pasquini}\ \emph {et~al.}(2014)\citenamefont
  {Pasquini}, \citenamefont {Polyakov},\ and\ \citenamefont
  {Vanderhaeghen}}]{PasquiniPolyakovVanderhaeghen2014}%
  \BibitemOpen
  \bibfield  {author} {\bibinfo {author} {\bibfnamefont {B.}~\bibnamefont
  {Pasquini}}, \bibinfo {author} {\bibfnamefont {M.~V.}\ \bibnamefont
  {Polyakov}},\ and\ \bibinfo {author} {\bibfnamefont {M.}~\bibnamefont
  {Vanderhaeghen}},\ }\bibfield  {title} {\bibinfo {title} {Dispersive
  evaluation of the {$D$}-term form factor in deeply virtual compton
  scattering},\ }\href {https://doi.org/10.1016/j.physletb.2014.10.047}
  {\bibfield  {journal} {\bibinfo  {journal} {Phys. Lett. B}\ }\textbf
  {\bibinfo {volume} {739}},\ \bibinfo {pages} {133} (\bibinfo {year}
  {2014})},\ \Eprint {https://arxiv.org/abs/1407.5960} {arXiv:1407.5960
  [hep-ph]} \BibitemShut {NoStop}%
\bibitem [{\citenamefont {Lorc{\'e}}\ \emph {et~al.}(2019)\citenamefont
  {Lorc{\'e}}, \citenamefont {Moutarde},\ and\ \citenamefont
  {Trawi{\'n}ski}}]{LorceMoutardeTrawinski2019}%
  \BibitemOpen
  \bibfield  {author} {\bibinfo {author} {\bibfnamefont {C.}~\bibnamefont
  {Lorc{\'e}}}, \bibinfo {author} {\bibfnamefont {H.}~\bibnamefont
  {Moutarde}},\ and\ \bibinfo {author} {\bibfnamefont {A.~P.}\ \bibnamefont
  {Trawi{\'n}ski}},\ }\bibfield  {title} {\bibinfo {title} {Revisiting the
  mechanical properties of the nucleon},\ }\href
  {https://doi.org/10.1140/epjc/s10052-019-6572-3} {\bibfield  {journal}
  {\bibinfo  {journal} {Eur. Phys. J. C}\ }\textbf {\bibinfo {volume} {79}},\
  \bibinfo {pages} {89} (\bibinfo {year} {2019})},\ \Eprint
  {https://arxiv.org/abs/1810.09837} {arXiv:1810.09837 [hep-ph]} \BibitemShut
  {NoStop}%
\end{thebibliography}%

\end{document}